\PassOptionsToPackage{dvipsnames, usenames}{xcolor}
\documentclass[twocolumn, twocolappendix]{aastex631}

\usepackage{acronym}
\usepackage{bold-extra}
\usepackage[T1]{fontenc}
\usepackage{multirow}
\usepackage{physics}
\usepackage{orcidlink}
\usepackage{xspace}
\usepackage{makecell}
\usepackage[normalem]{ulem}

\newacro{PE}[PE]{parameter estimation}
\newacro{GW}[GW]{gravitational-wave}
\newacro{GWTC-3}[GWTC-3]{third gravitational-wave transient catalog}
\newacro{GWTC-2}[GWTC-2]{second gravitational-wave transient catalog}
\newacro{GWTC}[GWTC]{gravitational wave transient catalog}
\newacro{LVK}[LVK]{LIGO-Virgo-Kagra collaboration}
\newacro{O4}[O4]{fourth observing run}
\newacro{O3}[O3]{third observing run}
\newacro{HMC}[HMC]{Hamiltonian Monte Carlo}
\newacro{MC}[MC]{Monte Carlo}
\newacro{NUTS}[NUTS]{No-U-Turn Sampler}
\newacro{IID}[IID]{identically and independently distributed}
\newacro{SFR}[SFR]{star formation rate}
\newacro{KDE}[KDE]{kernel density estimate}
\newacro{BBH}[BBH]{binary black-hole}
\newacro{BNS}[BNS]{binary neutron star}
\newacro{NSBH}[NSBH]{neutron star-black hole binary}
\newacro{NS}[NS]{neutron star}
\newacro{BH}[BH]{black hole}

\newacro{PPC}[PPC]{posterior predictive check}
\newacro{MI}[MI]{mutual information}
\newacro{CAR}[CAR]{conditional autoregressive}
\newacro{ICAR}[ICAR]{intrinsic conditional autoregressive}
\newacro{PDF}[PDF]{probability density function}
\newacro{KL}[KL]{Kullback--Leibler}
\newacro{JS}[JS]{Jensen--Shannon}

\newacro{XLA}[XLA]{Accelerated Linear Algebra}
\newacro{HLV}[HLV]{Hanford, Livingston and Virgo}
\newacro{SNR}[SNR]{signal-to-noise ratio}
\newacro{IID}[IID]{independent and identically distributed}
\newacro{PSD}[PSD]{power spectral density}
\newacro{DAG}[DAG]{directed acyclic graph}

\newacro{CE}[CE]{common-envelope}
\newacro{OC}[OC]{open cluster}
\newacro{NSC}[NSC]{nuclear star cluster}
\newacro{CHE}[CHE]{chemically-homogeneous evolution}
\newacro{GC}[GC]{globular cluster}
\newacro{AGN}[AGN]{active galactic nuclei}
\newacro{SMT}[SMT]{stable mass transfer}

\newacro{MS}[MS]{main sequence}

\newcommand{\xeffone}{0}

\newcommand{\zone}{0.2}
\newcommand{\ztwo}{0.8}

\newcommand{\chieff}{\ensuremath{\chi_\mathrm{eff}}}

\newcommand{\Msun}{M_\odot}
\newcommand{\rateunit}
{\mathrm{Gpc^{-3}\,yr^{-1}}}
\newcommand{\relrate}{\Delta\mathcal{R}/\bar{\mathcal{R}}}

\newcommand{\pr}{\textsc{Power Law}\xspace}
\newcommand{\PixelPop}{\textsc{PixelPop}\xspace}
\newcommand{\PeakTukey}{\textsc{Peak + Tukey}\xspace}
\newcommand{\Hybrid}{\textsc{Hybrid}\xspace}

\newcommand{\comment}[1]{}

\begin{document}

\title{
Nowhere left to hide: revealing realistic gravitational-wave populations in high dimensions and high resolution with \PixelPop}

\author{Sofía Álvarez-López\,\orcidlink{0009-0003-8040-4936}} 
\email{sofiaal@mit.edu}
\affiliation{LIGO Laboratory, Massachusetts Institute of Technology, Cambridge, MA 02139, USA}
\affiliation{Kavli Institute for Astrophysics and Space Research, Massachusetts Institute of Technology, Cambridge, MA 02139, USA}
\affiliation{Department of Physics, Massachusetts Institute of Technology, Cambridge, MA 02139, USA}

\author{Jack Heinzel\,\orcidlink{0000-0002-5794-821X}}      
\affiliation{LIGO Laboratory, Massachusetts Institute of Technology, Cambridge, MA 02139, USA}
\affiliation{Kavli Institute for Astrophysics and Space Research, Massachusetts Institute of Technology, Cambridge, MA 02139, USA}
\affiliation{Department of Physics, Massachusetts Institute of Technology, Cambridge, MA 02139, USA}

\author{Matthew Mould\,\orcidlink{0000-0001-5460-2910}}
\affiliation{LIGO Laboratory, Massachusetts Institute of Technology, Cambridge, MA 02139, USA}
\affiliation{Kavli Institute for Astrophysics and Space Research, Massachusetts Institute of Technology, Cambridge, MA 02139, USA}
\affiliation{Department of Physics, Massachusetts Institute of Technology, Cambridge, MA 02139, USA}
\affiliation{Nottingham Centre of Gravity \& School of Mathematical Sciences, University of Nottingham, University Park, Nottingham, NG7 2RD, United Kingdom}

\author{Salvatore Vitale\,\orcidlink{0000-0003-2700-0767}}  
\affiliation{LIGO Laboratory, Massachusetts Institute of Technology, Cambridge, MA 02139, USA}
\affiliation{Kavli Institute for Astrophysics and Space Research, Massachusetts Institute of Technology, Cambridge, MA 02139, USA}
\affiliation{Department of Physics, Massachusetts Institute of Technology, Cambridge, MA 02139, USA}

\begin{abstract}
The origins of merging compact binaries observed by the LIGO--Virgo--KAGRA gravitational-wave detectors remain uncertain, with multiple astrophysical channels possibly contributing to the merger rate. Formation processes can imprint nontrivial correlations in the underlying distribution of source properties, but current understanding of the overall population relies heavily on simplified and uncorrelated parametric models. In this work, we use \PixelPop---a high-resolution Bayesian nonparametric model with minimal assumptions---to analyze multidimensional correlations in the astrophysical distribution of masses, spins, and redshifts of black-hole mergers from mock gravitational-wave catalogs constructed using population-synthesis simulations. With full parameter estimation on 400 detections at current sensitivities, we show explicitly that neglecting population-level correlations biases inference. In contrast, modeling all significant correlations with \PixelPop allows us to correctly measure the astrophysical merger rate across all source parameters. We then propose a nonparametric method to distinguish between different formation channels by comparing the \PixelPop results back to astrophysical simulations. For our simulated catalog, we find that only formation channels with significantly different physical processes are distinguishable, whereas channels that share evolutionary stages are not. Given the substantial uncertainties in source formation, our results highlight the necessity of multidimensional astrophysics-agnostic models like \PixelPop for robust interpretation of gravitational-wave catalogs.
\end{abstract}

\section{Introduction}
\label{sec:introduction}

Up to the end of their \ac{O3}, the LIGO~\citep{TheLIGOScientific:2014jea}, Virgo~\citep{TheVirgo:2014hva}, and KAGRA~\citep{KAGRA:2020tym} (\acs{LVK}) \acf{GW} detectors have observed nearly 100 mergers of stellar-mass \acp{BH} and neutron stars~\citep{KAGRA:2021vkt}. While \acp{GW} encode information such as the distance, masses, and spins of the compact objects they originate from, analyzing many observations jointly allows us to infer the underlying astrophysical population~\citep{Vitale:2020aaz,Thrane:2018qnx,Callister:2024cdx,LIGOScientific:2018jsj,LIGOScientific:2020kqk,KAGRA:2021duu}. Broadly, the diverse range of astrophysical channels through which \ac{BBH} systems may form are grouped into two categories~\citep{Mapelli:2020vfa,Ivanova:2012vx,Bavera:2020uch,Gerosa:2021mno,Mapelli:2019bnp,Mandel:2021smh}: (a) isolated binary evolution and (b) formation in dynamical environments.

In scenario (a), \acp{BBH} form in isolation from the binary evolution of their massive stellar progenitors. Such systems may undergo \ac{SMT}~\citep{vandenHeuvel:2017pwp,Neijssel:2019irh,Gallegos-Garcia:2021hti,Briel:2022cfl} or \ac{CE} phases~\citep{1976IAUS...73.....E,Paczynski_1976,Tutukov:1993bse,Bethe:1998bn,2002ApJ...572..407B,Stevenson:2017tfq,Giacobbo:2018etu,Dominik:2012kk,Ivanova:2012vx,Bavera:2020uch}, which tighten the orbit sufficiently to enable a merger within a Hubble time~\citep{Peters:1964zz,Zevin:2020gbd}. Alternatively, if the progenitor stars are born in an already tight orbital configuration, tidal interactions can drive rapid stellar rotation, triggering \ac{CHE}~\citep{deMink:2009jq,Mandel:2015qlu, deMink:2016vkw, Marchant:2016wow,duBuisson:2020asn}. Despite only involving isolated binary evolution, \ac{CE} and \ac{SMT} share several evolutionary stages, while \ac{CHE} binaries undergo significantly different physical processes.

In scenario (b), stellar and BH binaries in dense environments such as \acp{GC}~\citep{Sigurdsson:1993zrm,Sigurdsson:1993tui,PortegiesZwart:1999nm,Miller:2002pg,Lightman:1978zz,Hut:1992wz,Fregeau:2006es,OLeary:2005vqo,Downing:2009ag,Rodriguez:2015oxa,Rodriguez:2016kxx,Rodriguez:2019huv,Askar:2016jwt}, \acp{NSC}~\citep{PortegiesZwart:1999nm,Miller:2008yw,Banerjee:2009hs,Rodriguez:2019huv,Zahn:1977mi}, and \ac{AGN} disks~\citep{Antonini:2016gqe,Mckernan:2017ssq,McKernan:2019beu,Stone:2016wzz,Tagawa:2019osr,Tagawa:2021ofj,Bartos_2017,Yang:2020lhq} undergo strong gravitational encounters that dynamically assemble merging binary systems. Hierarchical formation of \acp{BBH} in triple and multiple massive-star systems~\citep{Sana_2014,Silsbee:2016djf,Rodriguez:2018jqu,Gupta:2019unn,Toonen_2020} have also been predicted within these dense stellar environments.

To gain insight into the evolutionary channels shaping the population of binaries and infer the underlying distribution of \acp{BBH} from \ac{GW} data, Bayesian hierarchical inference is the state-of-the-art approach~\citep{Vitale:2020aaz,Thrane:2018qnx,Callister:2024cdx,LIGOScientific:2018jsj,LIGOScientific:2020kqk,KAGRA:2021duu}. This framework requires an input model for the underlying population, which is usually either parametric, astrophysically-informed, or flexible.

Heuristic parametric models describe the population using simple functional forms, perhaps with some astrophysical motivations. A few canonical examples include the \textsc{Power Law + Peak} mass model~\citep{Talbot:2018cva} and the \textsc{Power Law} redshift model~\citep{Fishbach:2018edt}. Parametric models also make strong assumptions about the underlying population, which may lead to model-driven conclusions, and their flexibility is limited by the functional forms they assume. As a result, they can introduce biases through model misspecification~\citep{Romero-Shaw:2022ctb,Cheng:2023ddt}.

Another approach is through astrophysically-informed or simulation-based models, which leverage astrophysical simulations as input to directly infer physically relevant quantities of stellar populations and binary evolution from \ac{GW} data~\citep{Barrett:2016edh, Taylor:2018iat, Wong:2020ise, Mould:2022ccw, Riley:2023jep, Leyde:2023iof, Mastrogiovanni:2022ykr, Zevin:2020gbd, Cheng:2023ddt, Colloms:2025hib, Neijssel:2019irh, Bouffanais:2020qds, Plunkett:2025mjr}. For instance, considering population synthesis simulations for five of the formation channels described above (\ac{CE}, \ac{SMT}, \ac{CHE}, \ac{GC}, and \ac{NSC}), \citet{Zevin:2020gbd} analyzed public \acs{LVK} data from the \ac{GWTC-2}. They found strong support for multiple evolutionary mechanisms in the underlying astrophysical population of detected \acp{BBH}, with the majority of sources inferred to evolve with a \ac{CE} episode, a preference for small natal \ac{BH} spins, and large  \ac{CE} efficiencies. These conclusions were supported by~\citet{Cheng:2023ddt} and~\citet{Colloms:2025hib} on \acs{GWTC-3} data. While simulation-based methods directly constrain the astrophysical parameters of these formation channels from \ac{GW} data, they make the strongest assumptions about uncertain physical processes. This means that considering an incomplete set of formation channels contributing to the underlying \ac{BBH} population or modeling the parameters with incorrect physical prescriptions can lead to significant biases~\citep{Cheng:2023ddt}. 

At the other end of the spectrum, nonparametric models offer the greatest flexibility by minimizing prior assumptions and avoiding restrictive constraints on the inferred population (see, e.g.,~\citet{Heinzel:2024jlc, Edelman:2022ydv, Godfrey:2023oxb, Golomb:2022bon, Heinzel:2023hlb, Tiwari:2020vym, Rinaldi:2021bhm, Callister:2023tgi, Li:2021ukd, Ray:2023upk, Farah:2024xub, Toubiana:2023egi,Sadiq:2023zee,Sadiq:2025aog,Sadiq:2025vly,Mandel:2016prl}). 
As a result, they are better able to uncover features that might otherwise be obscured by more rigid approaches, capturing the underlying astrophysical distribution of source properties with minimal bias. The downside is a bias--variance tradeoff, in that these models typically come with larger statistical uncertainties. Additionally, nonparametric models tend to be computationally expensive because---despite the name---they typically require inferring a larger number of parameters. Nonparametric models also lack interpretability due to their agnosticism about the underlying physics, although methods to reconstruct parametric models from nonparametric results have been proposed~\citep{Fabbri:2025faf,Rinaldi:2025evs}. 

Most nonparametric models are not designed to capture multidimensional correlations between source parameters, or to do so, must trade off resolution over the source parameter space due to the curse of dimensionality. Reducing the number of parameters can lower computational costs and allow for coarse-grained exploration of multidimensional structure. However, finer resolution is essential to capture the complex features and correlations that may be present in \ac{GW} data.
Failing to accurately model the population in multiple dimensions may obscure important astrophysical insights encoded in parameter correlations, which emerge naturally from the complex astrophysical processes governing formation channels~\citep{Bavera:2020uch,Bavera:2022mef,Gerosa:2021mno,Santini:2023ukl,Baibhav:2022qxm,vanSon:2021zpk,Zevin:2022wrw,Broekgaarden:2022nst,Bavera:2020inc}. Indeed, evidence for correlations has already been found in \ac{GW} data~\citep{Callister:2021fpo, Biscoveanu:2022qac, Adamcewicz:2022hce, Heinzel:2023hlb, Heinzel:2024hva, Franciolini:2022iaa, Ray:2024hos,Antonini:2024het,Hussain:2024qzl,Pierra:2024fbl,Tiwari:2021yvr,Li:2023yyt}, albeit mostly contingent on simplified population modeling assumptions.

To resolve these shortcomings, \citet{Heinzel:2024jlc, Heinzel:2024hva} introduced \PixelPop, a high-resolution nonparametric algorithm used to infer joint astrophysical distributions of---and multidimensional correlations between---the properties of GW sources, making minimal prior assumptions about the underlying population. In this paper, we use \PixelPop to assess the astrophysical conclusions that may be drawn using nonparametric Bayesian inference from GW data in a realistic setting. From \citet{Zevin:2020gbd}, we use the population-synthesis simulations of \ac{BBH} mergers that underwent \ac{CE} evolution to construct a GW catalog. For each event, we perform full Bayesian \ac{PE} at current detector sensitivities, as described in Sec.~\ref{sec:methods}. We pursue the following central questions:

\begin{enumerate}
\item \textit{\textbf{Can we capture the complex, high-dimensional correlations expected in realistic astrophysical populations?}} 

\PixelPop (described in Sec.~\ref{sec:pop inference}) allows us to recover the high-dimensional correlations arising from the formation of realistic \ac{GW} populations. However, doing so requires a highly sophisticated model that accounts for all significant correlations in the populations.

\item \textit{\textbf{What is the effect of neglecting correlations between source parameters?
}}

Neglecting correlations can lead to observable bias in the inferred population already with catalog sizes as expected by the end of the \ac{O4}~\citep{Capote:2024rmo} (Secs.~\ref{sec:2d-inference} and~\ref{sec:3d-inference}).
In contrast, modeling all relevant correlations enables correct inference of the astrophysical merger rate across all source parameters at the expense of increased computational cost and model complexity (Sec.~\ref{sec:(3+1)d-inference}).

\item \textit{\textbf{Can we extract astrophysical insights from nonparametric inference?
}}

We propose a similarity metric
(Sec.~\ref{sec:cross-entropy}) 
to distinguish between different formation channels by comparing them to the \PixelPop-inferred population.
Considering our \ac{O4}-like catalog, we find that a synthetic population consisting only of CE-driven BBH mergers is correctly favored over other formation channels with significantly different astrophysical processes (e.g., \ac{CHE}). However, a purely \ac{CE} population cannot be distinguished from a purely \ac{SMT} population,
likely due to the evolutionary stages shared between the two channels producing similar distributions of source properties.
\end{enumerate}

We summarize our methods and results before concluding with future applications of this work in Sec.~\ref{sec:conclusions}.

\section{Methods}
\label{sec:methods}

\subsection{Population-synthesis simulations}
\label{sec:pop synth simulations}

We consider a population of \ac{BBH} mergers that formed via \ac{CE} evolution---found to be the dominant channel by some analyses of current GW data with simulation-based models \citep{Zevin:2020gbd, Cheng:2023ddt,Colloms:2025hib}. We use simulations of binaries undergoing a late \ac{CE} phase by \citet{Zevin:2020gbd}, which are based on the \texttt{POSYDON} framework \citep{Fragos:2022jik} that integrates population synthesis from \texttt{COSMIC}~\citep{Breivik:2019lmt} with binary evolution calculations from \texttt{MESA}~\citep{Paxton:2010ji,Paxton:2013pj,Paxton:2015jva,Paxton:2017eie,Paxton:2019lxx}. To obtain the redshift distribution, \citet{Zevin:2020gbd} spread the binaries across cosmic time, assuming the star formation-rate density of~\citet{Madau:2016jbv} and a redshift-dependent metallicity prescription. The merger redshift is determined from the \ac{BBH} progenitor birth time, the \ac{BBH} formation time, and the binary inspiral time; see App. A.2 in~\citet{Zevin:2020gbd}. The resulting simulated population is parameterized by the natal BH spin $\chi_\mathrm{b}$ and the \ac{CE} ejection efficiency $\alpha_{\mathrm{CE}}$; for our population we take $\chi_\mathrm{b}=0$ and $\alpha_{\mathrm{CE}}=1$, following Table 1 in \citet{Cheng:2023ddt}. 

\begin{figure}
\centering
\includegraphics[width=\linewidth]{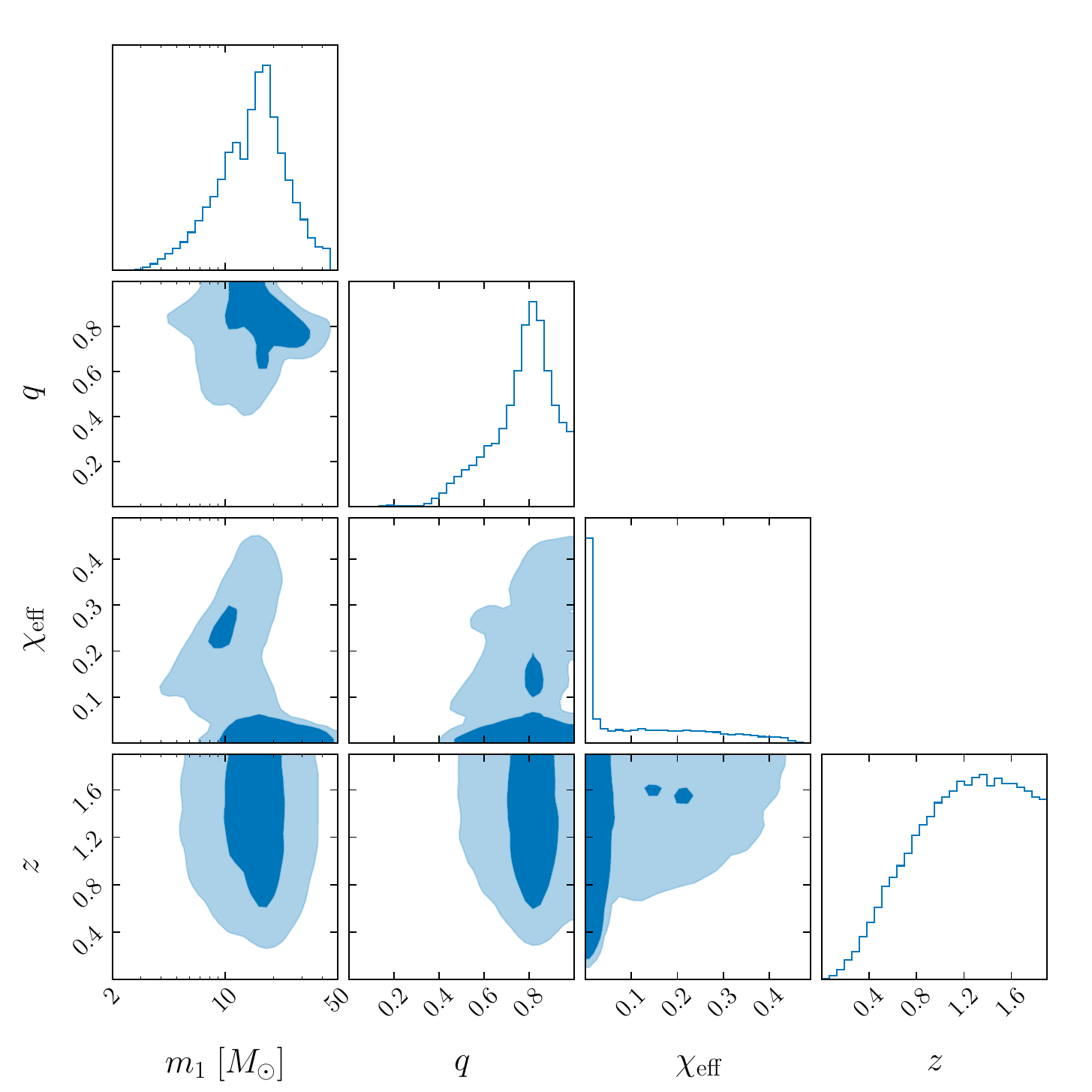}
\caption{
One- and two-dimensional marginals of the source-frame primary mass $m_1$, mass ratio $q$, effective spin $\chieff$, and redshift $z$ of the simulated \ac{CE} population. The contours show the 50\% and 90\% credible regions.
}
\label{fig:true_pop_corner_CE}
\end{figure}

We model the population in a four-dimensional space consisting of the BH source-frame primary mass $m_1$, binary mass ratio $q \in (0,1]$, redshift $z$, and effective aligned spin \citep{Racine:2008qv}
\begin{align}\label{eq:chi_eff}
    \chieff = \frac{a_1 \cos \theta_1 + qa_2 \cos \theta_2}{1 + q} \in (-1,1),
\end{align}
where $a_1$ and $a_2 \in [0, 1)$ are the dimensionless spin magnitudes, and $\theta_1$ and $\theta_2$ represent the spin--orbit misalignment of the primary and secondary components of the binary, respectively. We restrict the simulated population to $z\leq 1.9$, well above the highest redshift ($z=0.81$) of the detectable sources in our catalog (see Sec.~\ref{sec:detectable catalog} for details). Figure~\ref{fig:true_pop_corner_CE} shows the \ac{CE}-simulated population in the parameter space spanned by $m_1$, $q$, $\chieff$, and $z$. These four parameters are nontrivially and nonlinearly correlated. Typical population analyses with simplistic parameterizations would struggle to capture such complexities, motivating our use of multidimensional Bayesian nonparametrics like \textsc{PixelPop}. 

\subsection{Quantifying pairwise correlations in the simulated population}
\label{sec:MI}

To quantify the correlations in the simulated population (see Fig.~\ref{fig:true_pop_corner_CE}), we use the pairwise \ac{MI}~\citep{Kraskov:2004,Ross:2014}. The \ac{MI} is a nonnegative real number that quantifies the shared information between two random variables. In terms of the \ac{KL} divergence, the pairwise MI quantifies the information gained when considering the joint distribution of two random variables over their independent marginals; the higher the \ac{MI}, the stronger the correlation. The \ac{MI} between random variables $X$ and $Y$ is defined as~\citep{Laarne:2021}

\begin{align}
\label{eq: mutual information}
I(X,Y) =
\int \dd{x}\dd{y} p_{XY}(x,y)
\log \frac{ p_{XY}(x,y) }{ p_X(x) p_Y(y) }
\, ,
\end{align}
where $p_{XY}(x,y)$ is the joint \ac{PDF} for $X$ and $Y$, while $p_X(x)$ and $p_Y(y)$ are their univariate marginal distributions. In this paper, we use the \ac{MI} since the correlations in our astrophysical population are neither linear nor monotonic. Therefore, the covariances and (most) correlation coefficients (e.g., Pearson or Spearman) are not sufficient to capture the behavior of such correlations. To allow for easier interpretation, the \ac{MI} can be normalized to a correlation coefficient $\rho_I \in [0,1]$:
\begin{align}
\label{eq: mutual information coeff}
\rho_I(X,Y) \equiv \sqrt{1 - e^{-2I(X,Y)}},
\end{align}
such that $\rho_I(X,X) = \rho_I(Y,Y) = 1$ and $\rho_I(X,Y) = 0$ if and only if $X$ and $Y$ are independent. Note that this coefficient is an extension of the usual linear Pearson correlation coefficient~\citep{Laarne:2021}.

\begin{figure}
\includegraphics[width=\linewidth]{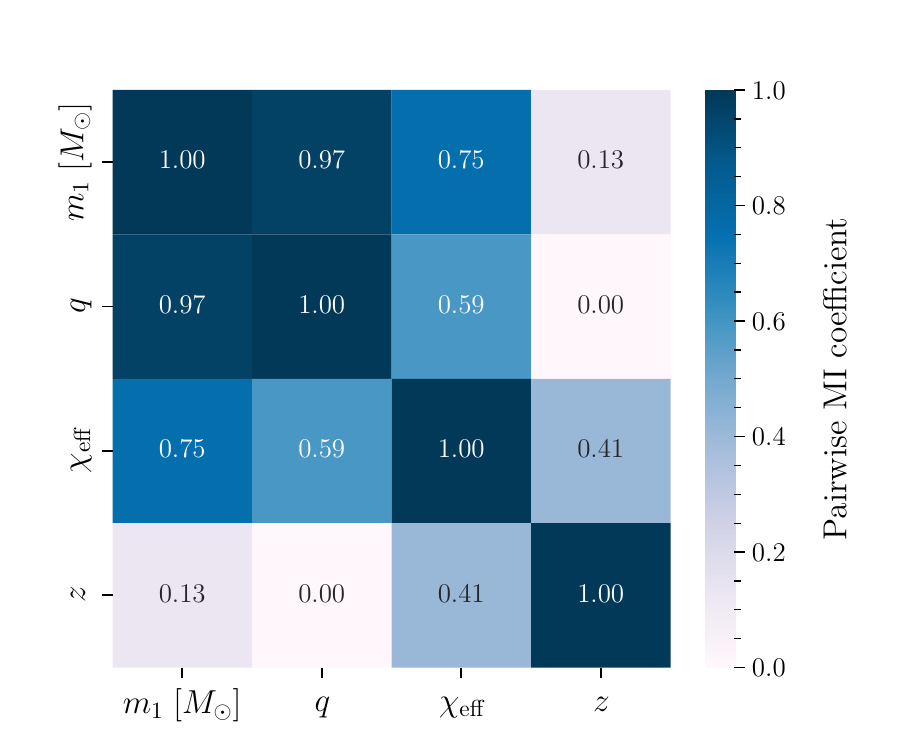}
\caption{Mutual-information correlation coefficient $\rho_I$ over the four-dimensional parameter space of primary mass $m_1$, mass ratio $q$, effective spin $\chieff$, and redshift $z$ for the \ac{CE} population.}
\label{fig:MI for CE pop}
\end{figure}

Figure~\ref{fig:MI for CE pop} shows the \ac{MI} correlation coefficient for all parameter pairs in the four-dimensional space spanned by $m_1$, $q$, $\chieff$, and $z$, calculated using the \texttt{ennemi} package~\citep{Laarne:2021}. The \ac{MI} coefficient indicates strong correlations between $m_1$ and $q$, with $\rho_I(m_1,q)\approx0.97$, as well as between $m_1$ and $\chieff$ with $\rho_I(m_1,\chieff)\approx0.75$ and between $q$ and $\chieff$ with $\rho_I(q,\chieff)\approx0.59$. We also find a moderate correlation between $\chieff$ and $z$: $\rho_I(\chieff,z)\approx0.41$. In contrast, note that $\rho_I(m_1,z)\approx0.13$ and $\rho_I(q,z) \approx 0$, which means that both primary mass and mass ratio are nearly independent of redshift in the simulated population. We note that other formation channels or modeling prescriptions may result in stronger correlations between mass and redshift~\citep[e.g.,][]{vanSon:2021zpk}. In Sec.~\ref{sec:results}, we use the \ac{MI} coefficients to progressively find the subset of \ac{BBH} source parameters with significant population-level correlations that must be captured with a multidimensional model to ensure unbiased inference.

\subsection{Building a catalog of detectable sources}\label{sec:detectable catalog}

We build a set of ${N_\mathrm{obs} = 400}$ \ac{GW} events (consistent with near-future catalogs \citep{Capote:2024rmo}) from the \ac{CE} population, considering a \ac{GW} interferometer network comprising both LIGO detectors and Virgo operating at O4 design sensitivity \citep{KAGRA:2013rdx, OReilly:2022}. We draw sources from the \ac{CE}-simulated population and calculate their optimal network \ac{SNR}. We consider a \ac{BBH} detected and add it to the catalog if $\rho_{\mathrm{opt}} \geq 11$. We note that the optimal \ac{SNR} implicitly depends on the true source parameters, so it is not a physical statistic on which to select detections~\citep{Essick:2023upv}. However, the resulting systematic bias is expected to be small and largely masked by statistical uncertainty for catalog sizes with $\mathcal{O}(100)$ events, like ours~\citep{Heinzel:2024jlc}.

For each of the \acp{BBH} in our catalog, we perform full \acf{PE} to draw posterior samples of the source parameters. We use the \texttt{IMRPhenomXP} waveform model \citep{Pratten:2020ceb} and relative binning~\citep{Cornish:2010kf, Cornish:2021lje, Zackay:2018qdy,krishna2023:bug}; as implemented in the Bayesian-inference software \texttt{Bilby}~\citep{Ashton:2018jfp} to accelerate likelihood evaluations. We set uniform priors on dimensionless spin magnitudes $\in[0,1)$, detector-frame chirp mass, and mass ratio $q$. We use isotropic priors for extrinsic parameters (e.g., sky location) and the spin tilts. Following~\citet{Cheng:2023ddt}, we adopt a uniform prior in redshift, rather than \texttt{Bilby}'s default (uniform in comoving volume per unit source-frame time). This prior choice does not affect our population inference, which is performed using the standard hierarchical framework described in Sec.~\ref{sec:pop inference}.

Applying a detection threshold to the simulated \acp{BBH} imposes a selection effect that we must account for~\citep{Thrane:2018qnx, Loredo:2004nn, Mandel:2018mve}. We therefore perform a simulation campaign to characterize the sensitivity of the simulated \ac{O4} LIGO--Virgo network to \acp{BBH} spanning a range of parameters, as described in App.~\ref{app:vtInjs}. We then estimate the selection efficiency using Monte Carlo integration~\citep{Tiwari:2017ndi,Farr_2019}.

\subsection{Population inference}
\label{sec:pop inference}

In this paper, we model the detection of \ac{GW} signals as an inhomogeneous Poisson process~\citep{Mandel:2018mve,Thrane:2018qnx,Vitale:2020aaz,Essick:2023upv}. We directly infer the \ac{BBH} merger rate $\mathcal{R}(\theta';z)$, that is, the number $N$ of mergers with source parameters $\theta'$ per unit comoving volume $V_\mathrm{c}$ and source-frame time $t_\mathrm{s}$:
\begin{align}
\label{eq:source-frame-rate}
\mathcal{R} ( \theta' ; z )
=
\frac
{ \dd{N} }
{ \dd{V_\mathrm{c}} \dd{t_\mathrm{s}} \dd{\theta'} }.
\end{align}
We emphasize that $\mathcal{R}(\theta';z)$ evolves with redshift $z$, but does not correspond to a density in $z$. We model $\mathcal{R}$ using \PixelPop, following the procedure described in Sections II and III of~\citet{Heinzel:2024jlc}. The only modification we introduce here is the use of an \ac{ICAR} prior, a limiting special case of the \ac{CAR} prior; we discuss the implemented changes in App.~\ref{sec:appICAR}.

With minimal astrophysical or parametric assumptions, \PixelPop discretizes the joint space spanned by the source parameters into uniformly spaced bins, $\theta_b$ ($b = 1,\ldots,B$), and infers the merger rate $\ln\mathcal{R}_b$ in each bin. The computational efficiency of \PixelPop allows us to employ a much higher binning resolution than, e.g., binned Gaussian processes (BGPs)~\citep{Ray:2023upk,KAGRA:2021duu}. We note that in our \PixelPop implementation (as in~\citealt{Heinzel:2024jlc}), neighboring bins are coupled with the same strength in all dimensions, whereas BGPs can infer a separate length scale per dimension. This is a modeling choice rather than an intrinsic constraint of the ICAR prior, and we do not find it to limit the analyses reported in this paper. The higher resolution nonetheless makes it possible to resolve finer details in the population and to model more dimensions.

\section{Inferring multidimensional astrophysical populations}
\label{sec:results}

\subsection{Are two-dimensional correlations sufficient?}
\label{sec:2d-inference}

Figure~\ref{fig:MI for CE pop} shows that $m_1$ and $q$ are the most strongly correlated parameters in the \ac{CE} population. Since studying different pairwise correlations is a common approach~\citep{Heinzel:2024hva}, we first use \PixelPop to infer the two-dimensional merger-rate density $\mathcal{R}(m_1,q)$. We use 100 bins per dimension, following \citet{Heinzel:2024jlc}. To model the remaining parameters, we leverage our knowledge of the true distributions (we discuss the implications of doing so in Sec.~\ref{sec:conclusions}). For redshift, we consider a \pr model~\citep{Fishbach:2018edt}. For the effective spin, we use a mixture model consisting of a narrow Gaussian to capture the sharp peak of the distribution (as seen in Fig.~\ref{fig:true_pop_corner_CE}) and a Tukey window spanning the range $0 < \chieff < 0.5$ to capture the bulk; see Eq.~(E.1) in \citet{Vitale:2022dpa}. We refer to this as the \PeakTukey model.
The parameters and priors we used for these parametric models can be found in Table~\ref{tab:parametric-params}. Note that the parametric models are independent from one another and from the merger rate over $m_1$ and $q$ given by \PixelPop. Overall, our model for the merger rate is given by:
\begin{align}
\mathcal{R} ( m_1 , q , \chi_\mathrm{eff} ; z )
=
\mathcal{R} ( m_1, q)
p ( \chieff ) (1+z)^{\kappa}
\, .
\label{eq: 2d rate}
\end{align}

\begin{table}
\centering
\renewcommand{\arraystretch}{1.15}
\setlength{\tabcolsep}{0pt}
\begin{tabular*}{0.95\linewidth}{@{\extracolsep{\fill}} l l c @{}}
\hline\hline
\noalign{\smallskip}
\textbf{Parameter} & \textbf{Description} & \textbf{Prior} \\
\noalign{\smallskip}
\hline
\noalign{\smallskip}
$\kappa$ & $z$ power-law index & $U(-2,5)$ \\
$\mu_{\chieff}$ & Narrow Gaussian location & $U(-1,0.05)$ \\
$\sigma_{\chieff}$ & Narrow Gaussian width & $U(0.0025,0.1)$ \\
$T_{x0}$ & Tukey window center & $U(0.05,1)$ \\
$T_k$ & Tukey window width & $U(0.1,1)$ \\
$T_r$ & Tukey window shape & $U(0,1)$ \\[4pt]
$\lambda_{\chieff}$ &
\shortstack[l]{Fraction of sources in the\\$\chi_{\mathrm{eff}}$ Gaussian component} &
$U(0,1)$ \\
\noalign{\smallskip}
\hline\hline
\end{tabular*}
\caption{Priors for the parameters of the \PeakTukey spin model and \pr redshift model. Uniform priors $U$ are used for all parameters. In spin, the priors are defined to avoid degeneracies between the narrow component peaking at $\chieff\approx0$ and the broader region of the distribution (see App. E in \citet{Vitale:2022dpa} for details on the Tukey window parameters).}
\label{tab:parametric-params}
\end{table}

\begin{figure*}
\centering
\includegraphics[width=\columnwidth]{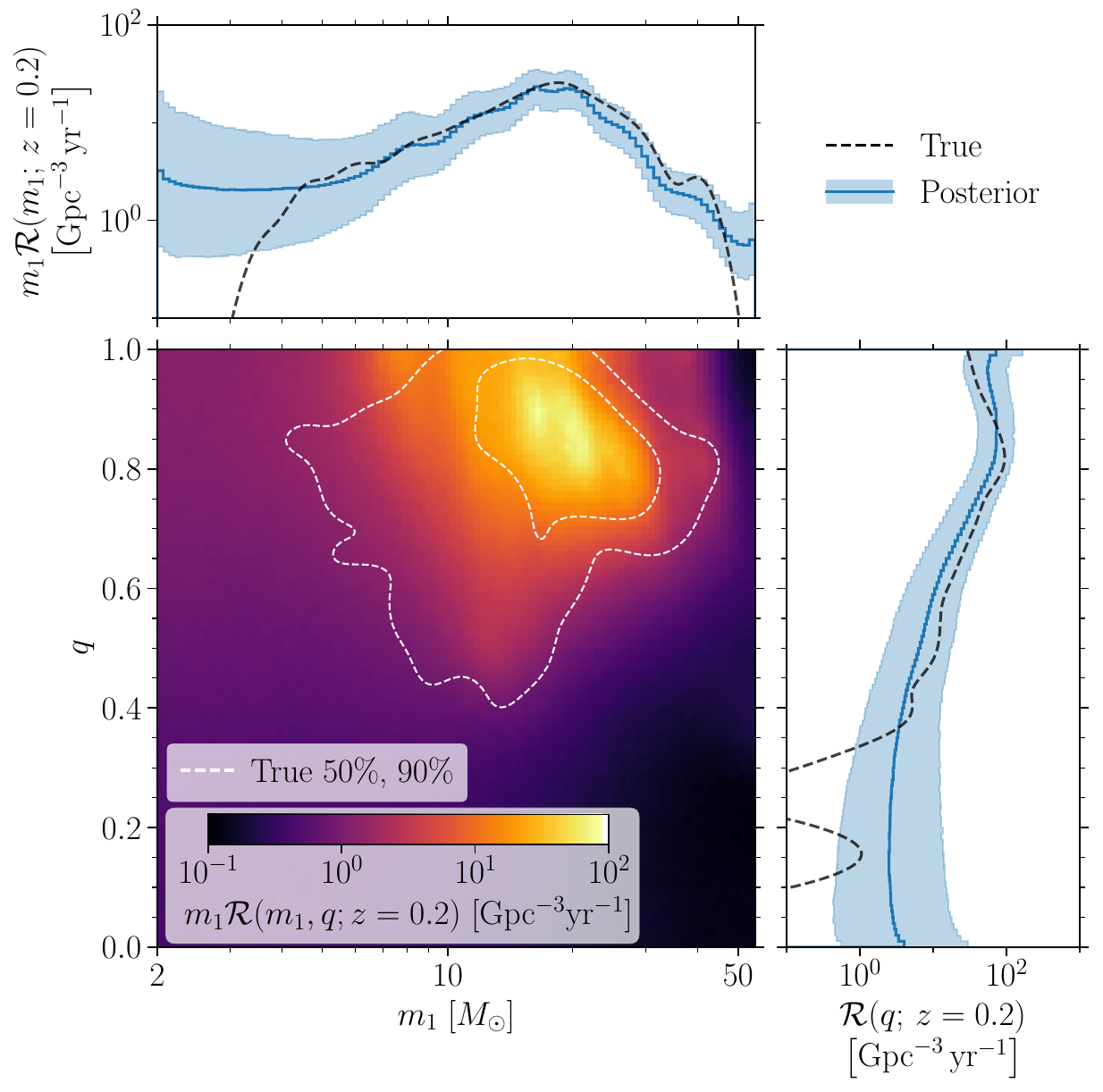}
\includegraphics[width=\columnwidth]{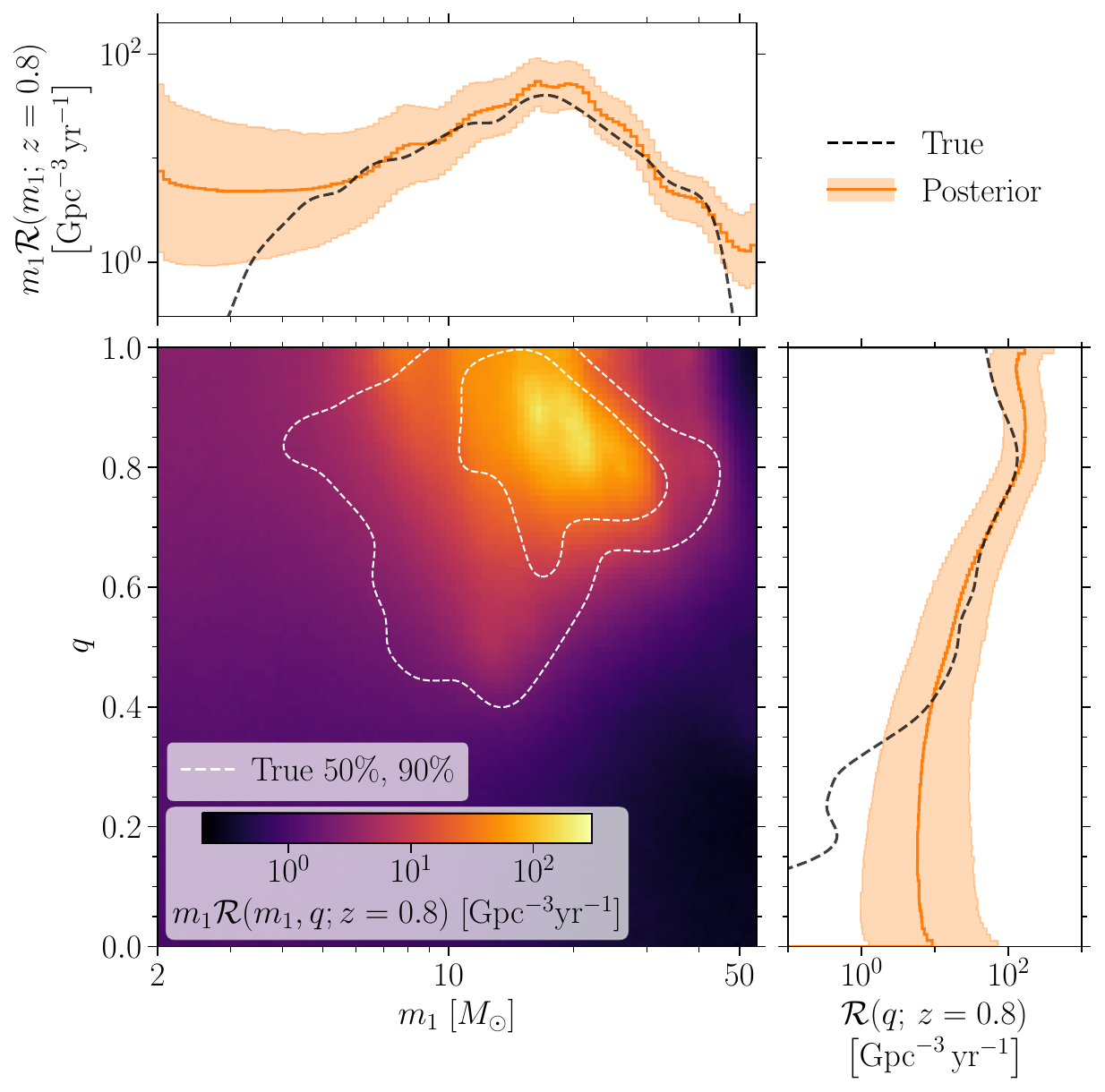}
\caption{
\textit{Left.} 
Comoving merger rate density $\mathcal{R}(m_1,q;z)$ inferred by \PixelPop, evaluated at a fixed redshift $z=0.2$. The central panel shows the median of the two-dimensional merger rate density, along with 50\% and 90\% credibility regions of the true population. Brighter regions indicate a higher merger rate. The upper and right-hand panels show the marginal merger rate density $\mathcal{R}(m_1;z=0.2)$ and $\mathcal{R}(q;z=0.2)$, respectively. The solid blue line shows the posterior median, with the central 90\% credible region shaded. The dashed lines show the true underlying population, calculated using a \acf{KDE} scaled by the overall merger rate implied by the number of detections during the observing time.
\textit{Right.} 
Same as the left-hand panel, except at $z = 0.8$. 
}
\label{fig:2d-z0p2p8}
\end{figure*}

On the left of Fig.~\ref{fig:2d-z0p2p8}, we show the inferred comoving merger-rate density $\mathcal{R}(m_1, q; z=0.2)$ evaluated at a redshift of $z=0.2$, marginalized over the effective-spin distribution. Note that \PixelPop successfully recovers the underlying $m_1$--$q$ correlation. Additionally, in the upper and right-hand panels, we present the one-dimensional marginal merger rate densities $\mathcal{R}(m_1;z=\zone)$ and $\mathcal{R}(q;z=\zone)$, respectively. For both $m_1$ and $q$, \PixelPop recovers the true distributions within the 90\% posterior uncertainty except in regions where we would not expect $\gtrsim\mathcal{O}(1)$ detection.

Note that our result deviates from the true population where the true merger rate is low (e.g., $m_1 \lesssim 3M_\odot$ and $q \lesssim 0.4$). This is characteristic of nonparametric models~\citep{Heinzel:2024jlc}: in particular, they overconfidently exclude merger-rates of $0\,\rateunit$ in the absence of informative data; e.g., in regions of low detectability (see discussion in Section IV of~\citet{Heinzel:2024jlc}). Additionally, note a slight increase in the inferred rate (and uncertainties) near the edges of parameter space (e.g., $q\approx1$). Bins at the boundary of parameter space are coupled to fewer neighboring bins, meaning the uncertainty is generically larger, but larger by a symmetric amount in $\ln\mathcal{R}$. When we marginalize over a range of bins, we integrate over $\mathcal{R}$ rather than $\ln\mathcal{R}$. The uncertainty thus becomes skewed to larger values at larger rates due to the nonlinearity of the exponential function.

\begin{table*}[t]
\centering
\setlength{\tabcolsep}{0pt}
\begin{tabular*}{\textwidth}{@{\extracolsep{\fill}} l c c c c @{}}
\hline\hline
\noalign{\smallskip}
\textbf{Model} & \multicolumn{2}{c}{\textbf{Primary Mass $\mathbf{m_1}$}} & \multicolumn{2}{c}{\textbf{Mass Ratio $q$}} \\
\noalign{\smallskip}
& \shortstack[c]{Smallest $\Delta\mathcal{R}/\bar{\mathcal{R}}$\\(location)}
& \shortstack[c]{Largest $\Delta\mathcal{R}/\bar{\mathcal{R}}$\\(location)}
& \shortstack[c]{Smallest $\Delta\mathcal{R}/\bar{\mathcal{R}}$\\(location)}
& \shortstack[c]{Largest $\Delta\mathcal{R}/\bar{\mathcal{R}}$\\(location)} \\
\noalign{\smallskip}
\hline
\noalign{\smallskip}
Two-dim. \PixelPop & 0.9 ($m_1\approx16\,\Msun$) & 3.3 ($m_1\approx4\,\Msun$) & 1.1 ($q\approx0.9$) & 2.9 ($q\approx0.4$) \\
Three-dim. \PixelPop & 0.8 ($m_1\approx21\,\Msun$) & 3.7 ($m_1\approx5\,\Msun$) & 1.2 ($q\approx0.9$) & 3.5 ($q\approx0.5$) \\
\Hybrid \PixelPop & 0.8 ($m_1\approx21\,\Msun$) & 3.1 ($m_1\approx4\,\Msun$) & 1.2 ($q\approx0.9$) & 2.9 ($q\approx0.4$) \\
\noalign{\smallskip}
\hline
\noalign{\smallskip}
Two-dim. \PixelPop on GWTC-3 & 3.2 ($m_1\approx33\,\Msun$) & 5.3 ($m_1\approx4\,\Msun$) & 3.9 ($q\approx0.8$) & 5.9 ($q\approx0.4$) \\
Fiducial parametric on GWTC-3 & 0.7 ($m_1\approx19\,\Msun$) & 1.8 ($m_1\approx40\,\Msun$) & 0.8 ($q\approx0.8$) & 1.9 ($q\approx0.4$) \\
\noalign{\smallskip}
\hline\hline
\end{tabular*}
\caption{Smallest and largest relative rate uncertainties $\relrate$, together with the parameter values at which they occur in parentheses, for primary mass and mass ratio, evaluated at $z=0.2$. Results are presented for the $m_1$ and $q$ marginals of two-dimensional \PixelPop (Sec.~\ref{sec:2d-inference}), three-dimensional \PixelPop (Sec.~\ref{sec:3d-inference}), and \Hybrid \PixelPop (Sec.~\ref{sec:(3+1)d-inference}) models on the parameters of the simulated \ac{CE} population. For reference, we also show the corresponding relative rate uncertainties obtained with the two-dimensional $m_1$--$q$ \PixelPop study of GWTC-3~\citep{Heinzel:2024hva} and with the fiducial parametric \textsc{Powerlaw+Peak} population model used in GWTC-3~\citep{KAGRA:2021duu}.}
\label{tab:relrates}
\end{table*}

Excluding prior-dominated regions, we quantify the precision of our inferred rates using the relative rates uncertainty given by the width of the central 90\% credible interval $\Delta\mathcal{R}$---within which we recover the true distributions---divided by the median inferred rate $\bar{\mathcal{R}}$. The smallest (largest) of these uncertainties corresponds to the most (least) precise measurement of the rate. Based on Fig.~\ref{fig:2d-z0p2p8}, we select $4\;\Msun \lesssim m_1 \lesssim 50 \;\Msun$ and $0.4 \lesssim q \leq 1$ as the non-prior-dominated regions. In the case of primary mass, the rate with smallest relative uncertainty is $23^{+12}_{-8}\,\rateunit$, near $m_1\approx16\,\Msun$. The width of the 90\% posterior credible interval is thus $\Delta\mathcal{R}=20\,\rateunit$, such that the smallest relative rate uncertainty is $\relrate=0.9$. We report this value in Table~\ref{tab:relrates}, along with the largest relative rate uncertainty in $m_1$ (and the value at which it occurs). We also include the smallest and largest $\relrate$ (and locations) for mass ratio. To put these values in context, we compare our measurements to the relative rate uncertainties in \acs{GWTC-3} using (a) two-dimensional $m_1$--$q$ \PixelPop as in~\citet{Heinzel:2024hva} and (b) the fiducial \textsc{Power Law + Peak} parametric model from~\citet{KAGRA:2021duu} (fourth and fifth rows in Table~\ref{tab:relrates}). Note that our relative uncertainties are narrower than those of model (a). In fact, our smallest values of $\relrate$ are similar to those of model (b). This is because we consider a catalog of 400 sources, comparable to the end of \ac{O4}, which is almost six times larger than \acs{GWTC-3}. However, our largest relative uncertainties are wider than in model (a) due to \PixelPop being a more flexible model.

Next we evaluate the merger rate at $z=\ztwo$, i.e., we consider $\mathcal{R}(m_1,q;z=\ztwo)$. On the right panel in Fig.~\ref{fig:2d-z0p2p8}, note that the inferred rate remains consistent with the true rate, even though we do not model the redshift evolution with \PixelPop. The fact that we still recover the true distributions for $m_1$ and $q$ suggests that these two variables are nearly independent of $z$. This behavior is consistent with the \ac{MI} coefficients we calculated in Fig.~\ref{fig:MI for CE pop}: $\rho_I(m_1,z) \approx 0.13$ and $\rho_I(q,z) \approx 0$, implying that the $m_1$--$z$ and $q$--$z$ correlations are negligible for the \ac{CE} population.

\begin{figure}
\centering
\includegraphics[width=\linewidth]{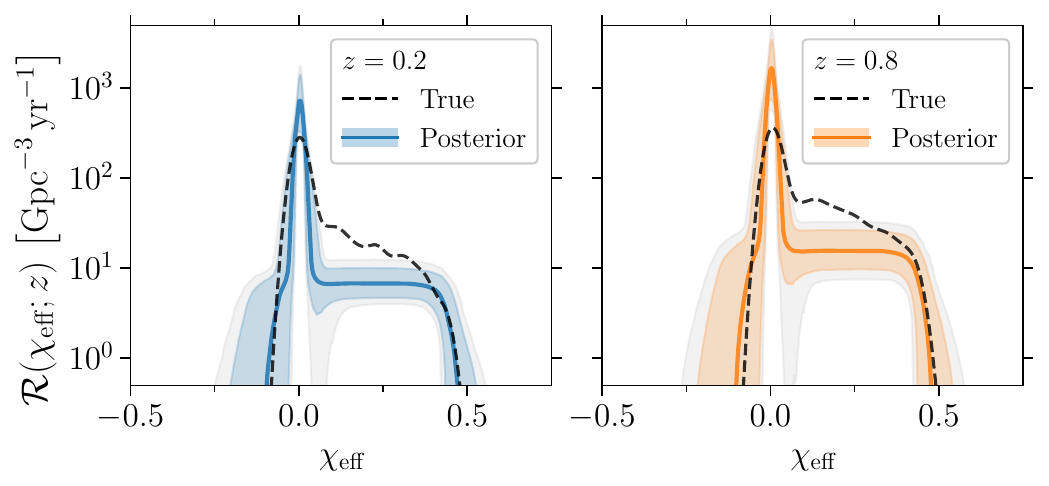}
\caption{Inferred merger-rate density $\mathcal{R}(\chieff; z)$ evaluated at redshifts $z=0.2$ (left, in blue) and $z=0.8$ (right, in orange). The solid line shows the posterior median, while the colored (gray) regions enclose the central 90\% (99\%) credible intervals. The dashed lines represent scaled \acp{KDE} of the true underlying population. }
\label{fig:2d-xeff-comparison}
\end{figure}

While the two-dimensional $m_1$--$q$ \PixelPop results appear unbiased, the parametric assumptions of Eq.~\eqref{eq: 2d rate} lead to major biases from the true population in effective spin. Using the \PeakTukey parameterization, we do not correctly recover the true underlying distribution, as shown in Fig.~\ref{fig:2d-xeff-comparison}. At both $z=\zone$ and $z=\ztwo$, the rate inferred with the \textsc{Peak + Tukey} model for $0 < \chieff < 0.5$ is below that of the true population, while the inferred rate for $\chieff \approx \xeffone$ is overestimated. At the lower redshift value, we find that the true peak of the distribution (i.e., $\chieff\approx\xeffone$) lies at the 2\% level of the inferred merger rate. Additionally, in the bulk of the distribution, the largest deviation from the true merger rate occurs at $\chi_\mathrm{eff} \approx 0.2$, where the inferred merger rate is inconsistent with the truth. The inconsistency between the true and inferred merger rates worsens at $z=0.8$, both at $\chieff\approx0$ and $\chieff\approx0.2$.

The fact that we do not recover the underlying $\chieff$ distribution is not due to the specific choice of \PeakTukey model. Rather, the bias arises from neglecting the $m_1$--$\chieff$ and $q$--$\chieff$ correlations in Eq.~\eqref{eq: 2d rate}, which are significant, as illustrated by the \ac{MI} coefficients in Fig.~\ref{fig:MI for CE pop}. To verify that this is the source of bias, in App.~\ref{sec:appUncorrPop} we artificially remove the correlations between the effective spin and the rest of the parameters and show that our parametric model is able to recover the behavior of the distribution with the 99\% credible intervals. Therefore, we find that a two-dimensional \PixelPop model is not sufficient and we need at least a three-dimensional model to capture the correlations in the simulated population.

\subsection{Are three-dimensional correlations sufficient?}
\label{sec:3d-inference}

In the interest of capturing the three strongest correlations in the \ac{CE} population, we now jointly model primary mass $m_1$, mass ratio $q$, and effective spin $\chi_\mathrm{eff}$ with \PixelPop.
By including $\chieff$, we account for the $m_1$--$\chieff$ and $q$--$\chieff$ correlations, which have the highest \ac{MI} coefficients after the $m_1$--$q$ pair (see Fig.~\ref{fig:MI for CE pop}). Our bins are now three-dimensional (i.e., voxels). As $10^6$ bins (100 bins per dimension) is computationally prohibitive, we reduce our binning resolution to 45 bins in each dimension, totaling $45^3 \approx 9\times10^4$ bins. Though decreasing the number of bins comes at the expense of reduced resolution, we found the above choice sufficient to resolve the joint and marginal distributions of the underlying population, as seen in the following. For redshift $z$, which is not modeled by \PixelPop, we again use the \pr redshift model, with priors in Table~\ref{tab:parametric-params}. The overall model for the merger-rate density is now
\begin{align}
\label{eq: 3d model}
\mathcal{R}(m_1,q,\chieff;z) = \mathcal{R}(m_1,q,\chieff) (1+z)^{\kappa} .
\end{align}
To visualize the three-dimensional results, we calculate the two-dimensional marginal rates of each parameter pair, as well as the one-dimensional marginals.

\begin{figure*}
\centering
\includegraphics[width=0.32\textwidth]{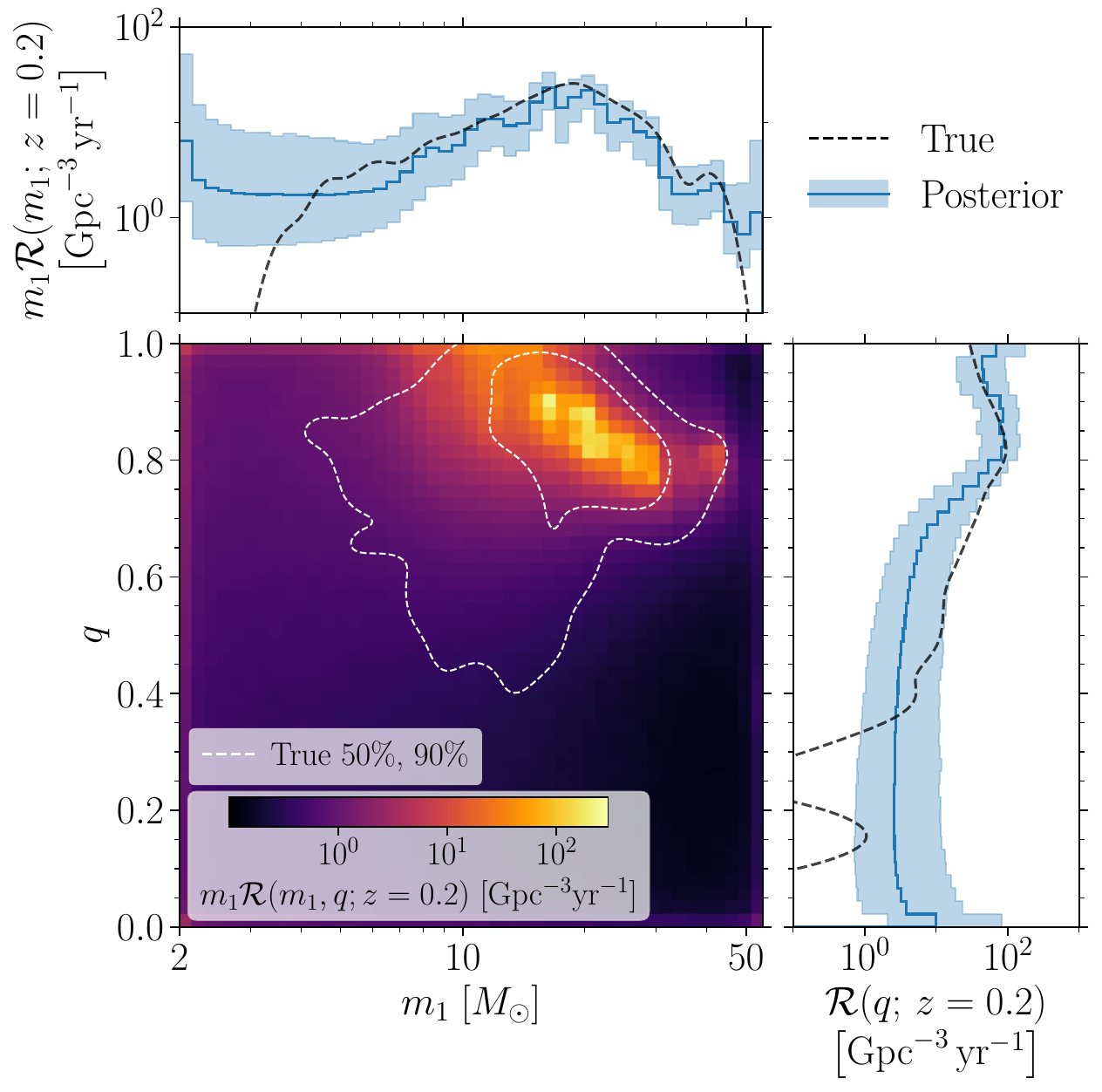}
\includegraphics[width=0.32\textwidth]{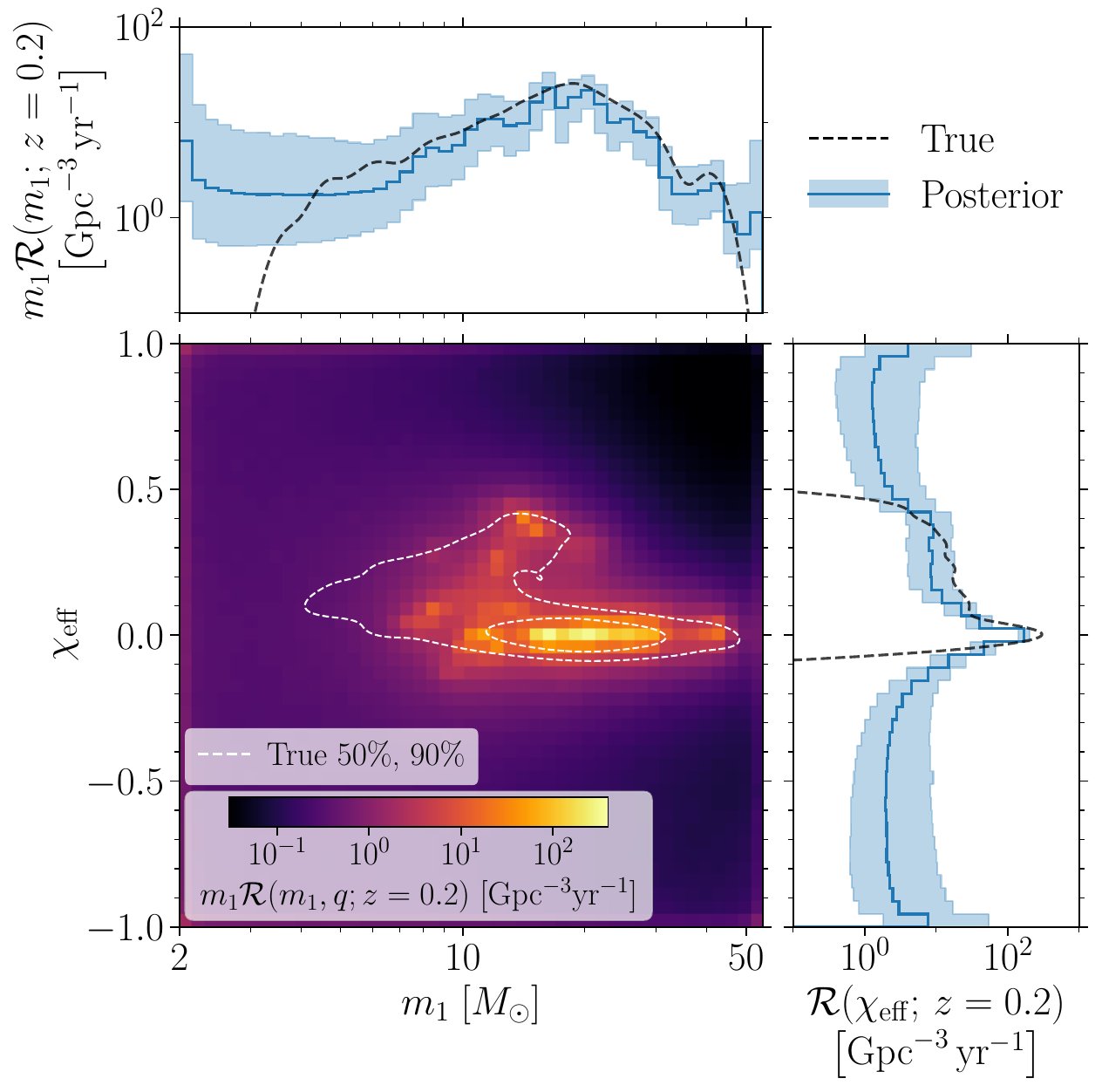}
\includegraphics[width=0.32\textwidth]{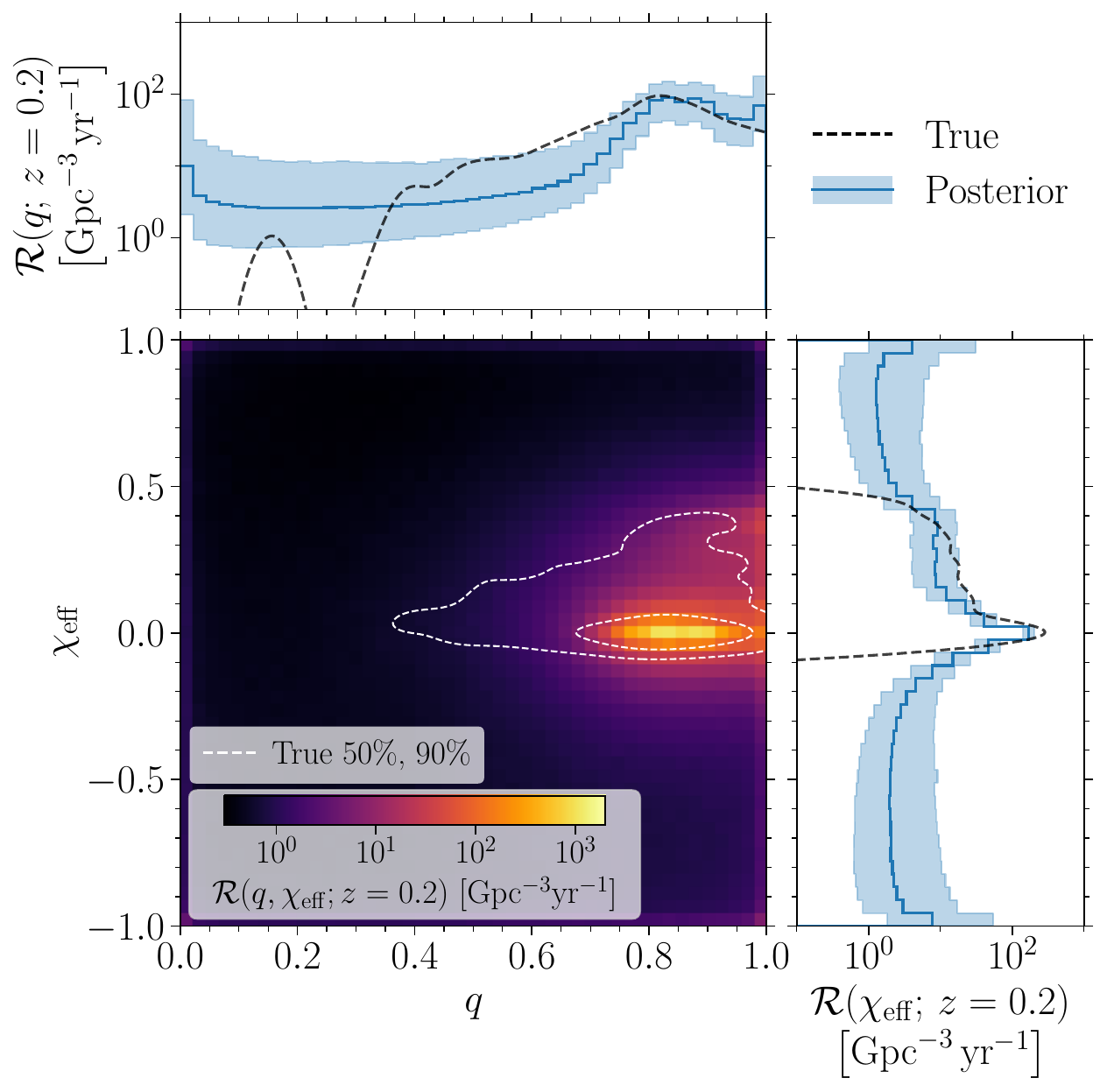}
\includegraphics[width=0.32\textwidth]{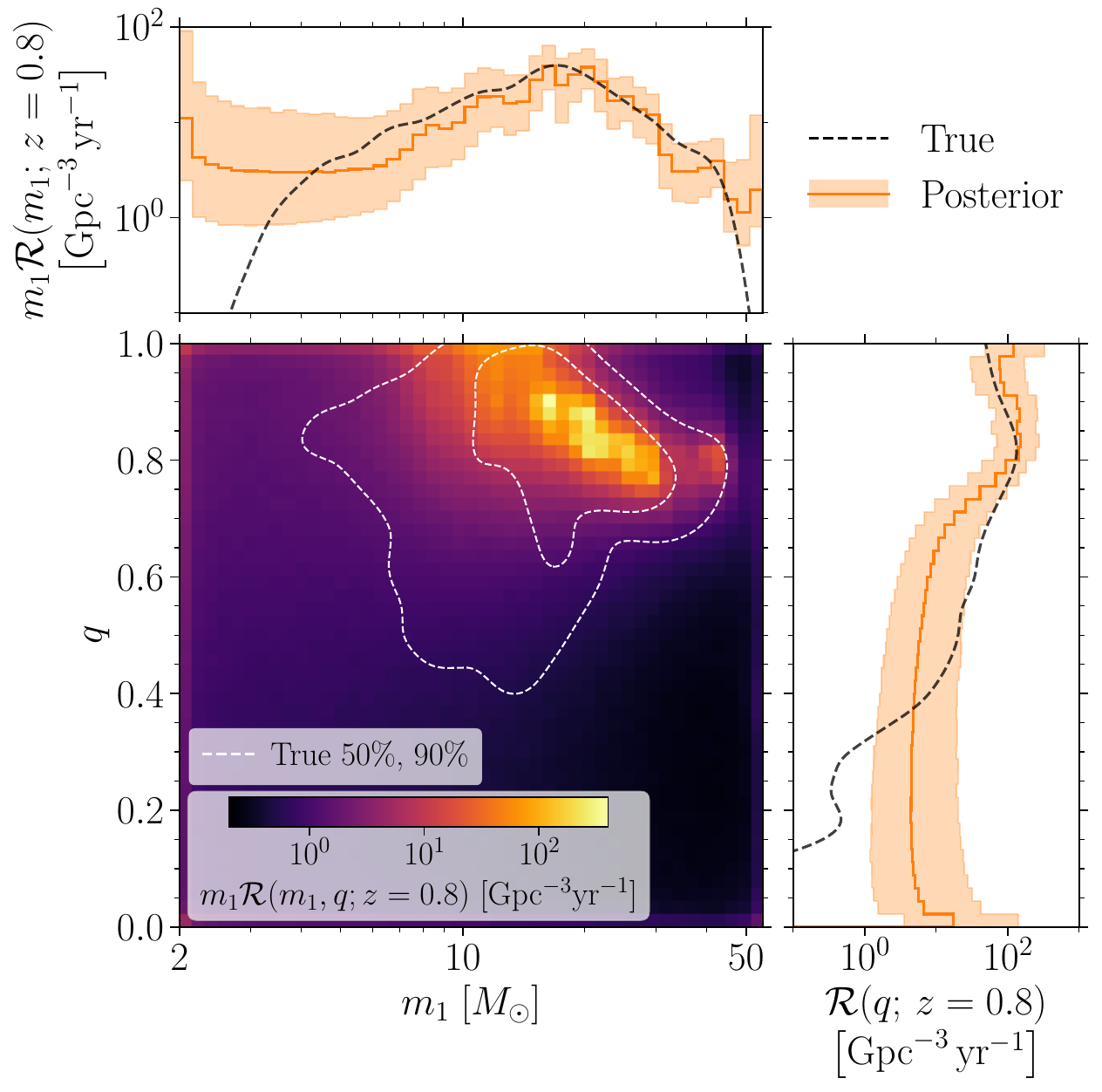}
\includegraphics[width=0.32\textwidth]{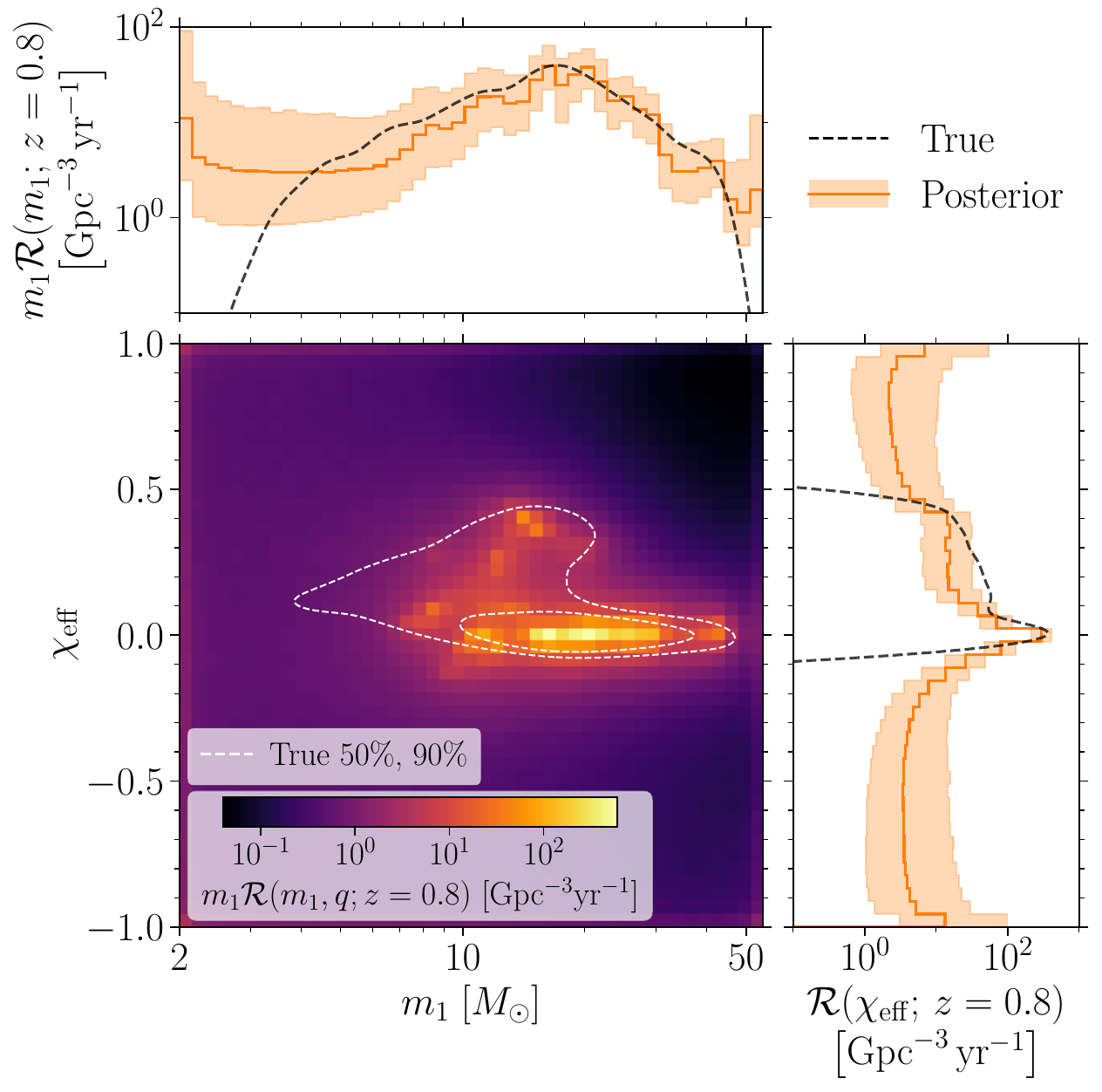}
\includegraphics[width=0.32\textwidth]{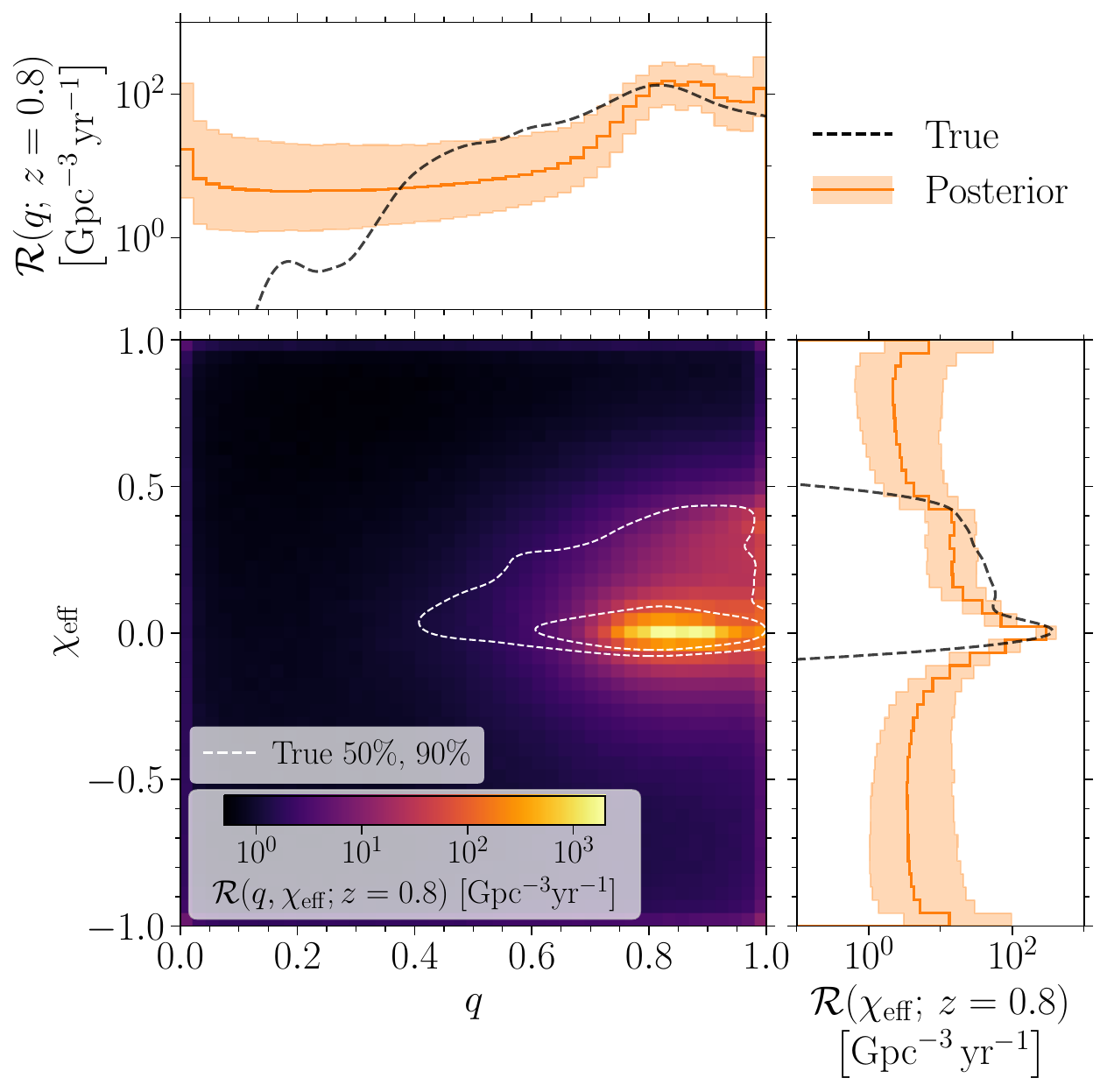}
\caption{
For our three-dimensional \PixelPop model, the inferred comoving merger-rate density $\mathcal{R}(m_1,q;z)$ (left), $\mathcal{R}(m_1,\chieff;z)$ (center), and $\mathcal{R}(q,\chieff;z)$ (right), evaluated at fixed redshifts $z=\zone$ (top) and $z=\ztwo$ (bottom). In each figure, the central panel shows the median of the two-dimensional merger-rate posterior, along with 50\% and 90\% credibility contours of the true underlying population of our simulated catalog; brighter regions indicate a higher inferred merger rate. Similarly, each upper (right-hand) panel shows the rate marginalized over the source parameter on the vertical (horizontal) axis; the dashed line represents the true population, while the solid line and shaded region represent the posterior median and central 90\% credible region.
}
\label{fig:3d}
\end{figure*}

Figure~\ref{fig:3d} shows the one- and two-dimensional marginals for the inferred merger-rate $\mathcal{R}(m_1, q,\chieff;z)$, evaluated at $z=\zone$ (top row) and $z=\ztwo$ (bottom row). \PixelPop recovers the true $m_1$ and $q$ distributions within the posterior uncertainty, except again in regions where we would not expect $\gtrsim\mathcal{O}(1)$ detection. The second row in Table~\ref{tab:relrates} shows the smallest and largest relative rate uncertainties we find for primary mass and mass ratio evaluated at $z=0.2$. These $\relrate$ values are comparable or marginally larger than those obtained with two-dimensional \PixelPop, consistent with the greater flexibility of the three-dimensional model.

At both $z=0.2$ and $z=0.8$, the inferred 90\% credible interval successfully recovers the true mass ratio distribution in most of the $0 < q \leq 1$ range, excluding prior-dominated regions. However, we see small deviations from the 90\% credible region for $0.5 \lesssim q \lesssim 0.7$. For instance, at the point of maximum deviation between the inferred and true rate, $q\approx0.65$, we find that the truth lies at the 97\% level of the inferred posterior distribution. In App.~\ref{app:appDeltas}, we show that these deviations are not attributable to model systematics. There, we run the analysis using point estimates for the properties of each event in our simulated catalog (i.e., in the absence of \ac{PE} uncertainty) and recover the true mass-ratio distribution within the 90\% uncertainty from \PixelPop (see Fig.~\ref{fig:q-comp} in App.~\ref{app:appDeltas}). 

At $z=0.2$, the true effective spin distribution in the range $0 < \chieff \leq 0.5$ is now enclosed within the 90\% credible interval inferred by \PixelPop, which constitutes a significant improvement from the distribution obtained with the parametric \PeakTukey model. We find that the smallest relative rate uncertainty in non-prior-dominated regions is $\relrate=0.4$ at the peak of the distribution and the largest is $\relrate=1.6$ near $\chieff\approx0.3$. 
 
\begin{figure}
\centering
\includegraphics[width=\linewidth]{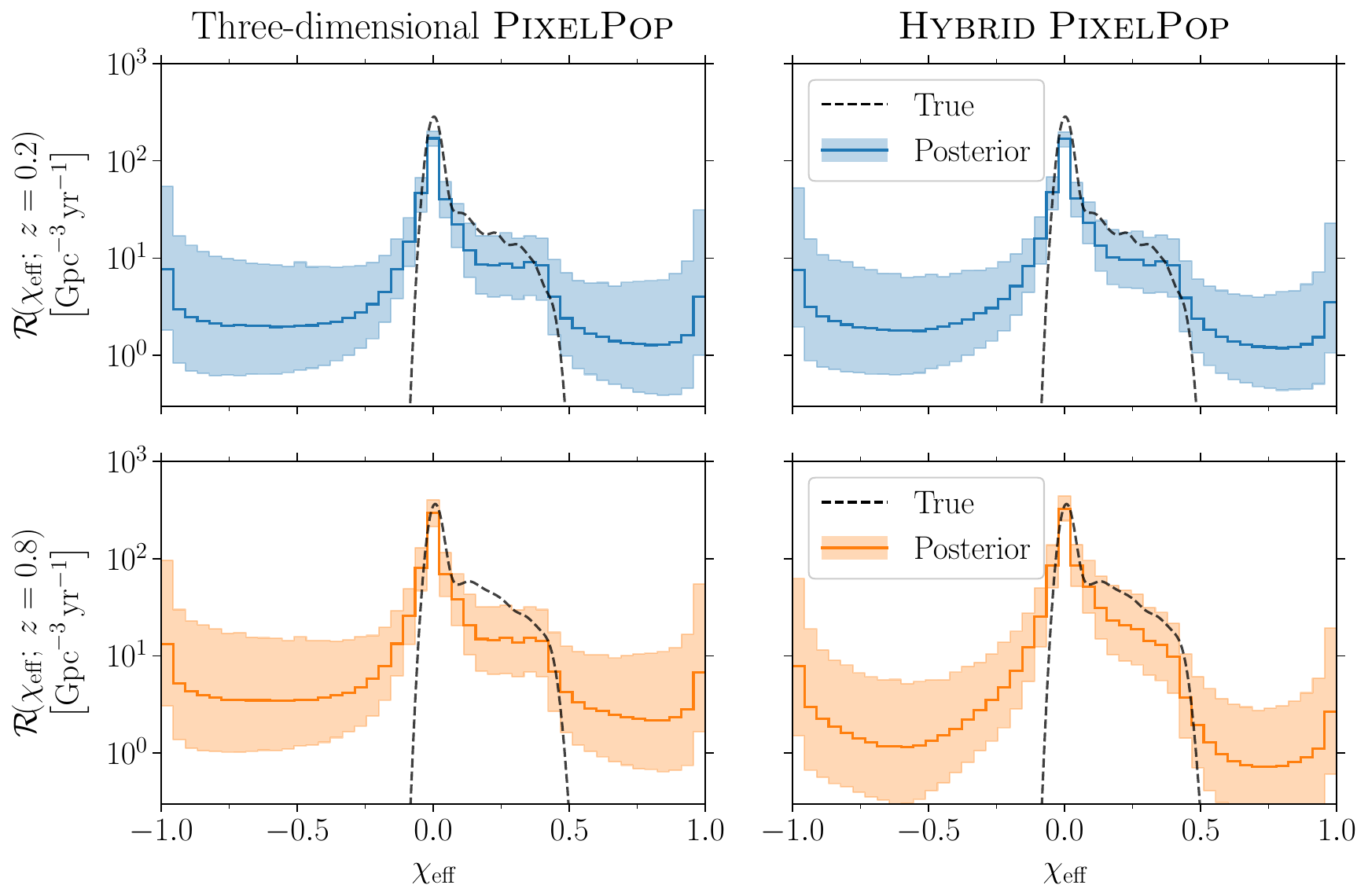}
\caption{Inferred merger-rate density $\mathcal{R}(\chieff ; z)$ evaluated at redshifts $z=0.2$ (blue) and $z=0.8$ (orange) for three-dimensional \PixelPop (left) and the \Hybrid model (right), defined in Eqs.~\eqref{eq: 3d model} and~\eqref{eq:(3+1)d-model}, respectively. The dashed line shows the true population, while the solid line and shaded region represent the posterior median and central 90\% credible region.
}
\label{fig:3d-comparison}
\end{figure}

However, at $\chieff\approx0$, the merger rate inferred by \PixelPop is lower than that of the true population. In fact, the true rate lies at the 99.9\% level of the inferred posterior distribution. This effect is clearer in the top-left panel of Fig.~\ref{fig:3d-comparison}, which compares $\mathcal{R}(\chieff;z)$ at $z=0.2$ (top) and $z=0.8$ (bottom). In App.~\ref{app:appDeltas}, we show that this bias at the peak is due to the imperfect approximation of the likelihood for narrow populations. However, it is worth noting that the sharpness of the peak in the true distribution likely arises from assumptions in the underlying astrophysics~\citep{Fuller:2019sxi}, which might not reflect a true population, as discussed in~\citet{Colloms:2025hib}.

We note that the \PeakTukey model introduced in Sec.~\ref{sec:2d-inference}, recovers the negative side of the $\chieff$ distribution and the region where $\chieff > 0.5$ better than \PixelPop. We stress that this is because we chose the parametric model to match the known truth from the \ac{CE} population synthesis simulations. In a realistic case, we do not know the correct functional form \textit{a priori}, which might lead to observable bias and severe model misspecification.

It is worth discussing the behavior of \PixelPop in regions of parameter space where few or no sources are detected. For example, for $\chieff < 0$ and $\chieff > 0.5$, there are no sources from the true population, so \PixelPop is only able to place an upper bound on the merger rate in these regions. Ideally, a model should infer a merger rate consistent with zero in regions without any sources. However, as discussed before, it is challenging for nonparametric models to measure low merger rates. Note, however, that in the three bins to the left of the $\chieff$ peak, i.e. for $-0.15 \leq \chieff \leq 0$, the measured merger rate exceeds $\mathcal{O}(1)\,\rateunit$. This is a result of the underlying assumptions of the \PixelPop prior. In particular, \PixelPop assumes a smooth coupling in log-space between nearest neighbors~\citep{Heinzel:2024jlc}, which makes it difficult to infer sharp drops in the merger rate. Although this smoothing may suppress some sharp features, omitting it would produce unreasonably noisy distributions. A further discussion of the challenges of \PixelPop and nonparametric models can be found in Sec. IV of~\citet{Heinzel:2024jlc}.  

\begin{figure}
\centering
\includegraphics[width=0.8\linewidth]{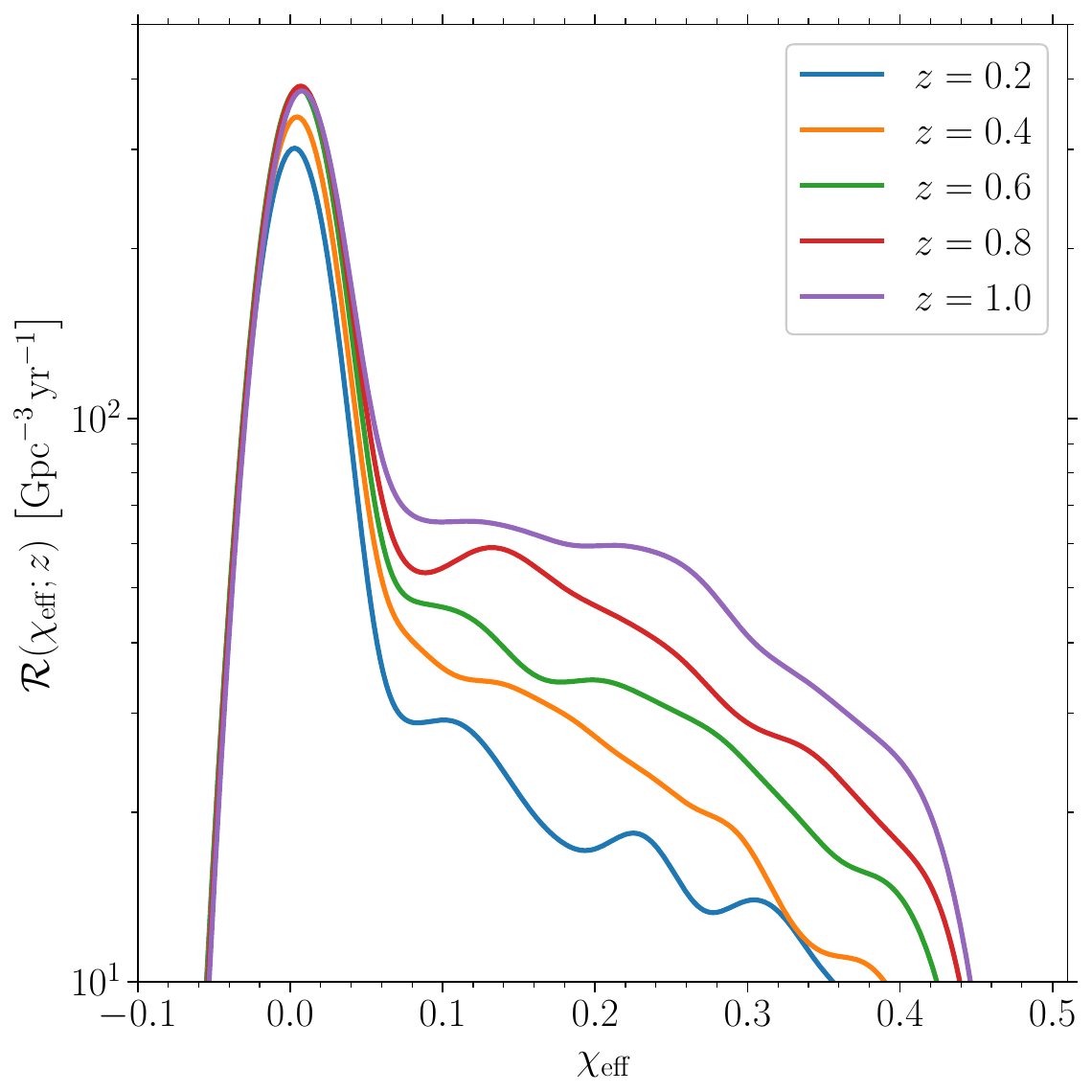}
\caption{Scaled \acp{KDE} of the merger rate $\mathcal{R}(\chieff;z)$ ias a function of effective spin $\chieff$, evaluated at redshifts $z \in \{0.2, 0.4, 0.6, 0.8, 1.0\}$. The shape of the distribution evolves with redshift.}
\label{fig:xeff-evolves-with-redshift}
\end{figure}

At $z=0.8$, the true value at the peak of the effective spin is now recovered within \PixelPop's uncertainty range: the true rate lies at the 87\% level of the inferred merger rate posterior distribution. This is because the $\chieff$ distribution at $z=\ztwo$ is slightly less steep from peak to trough than at $z=\zone$, as shown in Fig.~\ref{fig:xeff-evolves-with-redshift}. In fact, the peak--trough rate ratio at $z=\ztwo$ is $\sim 0.5$ times smaller than that at $z=\zone$. 
However, for $0 < \chieff \leq 0.5$, the marginal $\chieff$ distribution at $z=0.8$ deviates from the truth. This effect is more clearly seen in the bottom-left panels of Fig.~\ref{fig:3d-comparison}. In fact, at $\chieff\approx0.2$---the point of maximum deviation between the inferred and true rate---we find that the truth lies at the 99\% level of the inferred posterior distribution. As we show in App.~\ref{app:appDeltas}, when repeating the analysis using point estimates for the properties of each event, we \textit{still} find that the true merger rate at $\chieff\approx0.2$ lies at the 99\% level of the inferred posterior distribution. The fact that the true $\chieff$ rate is recovered at lower (but not higher) values of redshift is a symptom of ignoring the $\chieff$--$z$ correlation. Indeed, the shape of the $\chieff$ merger rate density evolves with redshift, as shown in Fig.~\ref{fig:xeff-evolves-with-redshift}. This is also consistent with the non-negligible \ac{MI} coefficient $\rho_I(\chieff,z)$ from Fig.~\ref{fig:MI for CE pop}. Therefore, we must model the $\chieff$--$z$ correlation.

\subsection{Are all correlations necessary?}
\label{sec:(3+1)d-inference}

The natural next step would be to model all source parameters with \PixelPop, which would allow us to effectively consider all correlations in our $(m_1,q,\chieff,z)$ parameter space. However, that is currently impractical for two main reasons. First, if we kept the same resolution as in three dimensions, in four dimensions we would now have $45$ times more bins, i.e., $45^4 \approx 4.1 \times 10^6$ total bins. Additionally, we find that the approximations used to compute the population likelihood (see e.g., Refs.~\citep{Vitale:2020aaz,Thrane:2018qnx,Callister:2024cdx,LIGOScientific:2018jsj,LIGOScientific:2020kqk,KAGRA:2021duu,Tiwari:2017ndi,Farr_2019}) break down dramatically for four-dimensional \PixelPop. With so many available pixels, it is inevitable that there will be some that contain one or more posterior samples from one of the observations but contain \textit{no} samples for estimating the selection efficiency. In that scenario, the likelihood estimator becomes unbounded, and the inferred rate in those bins diverges. It is possible that different approaches for evaluating the likelihood might be devised to avoid this issue, which we are exploring in ongoing work. An obvious solution that allows us to keep the same algorithm is to reduce the number of bins per dimension. We have decided not to follow that route because using fewer bins can wash out finer features in the population, e.g., the sharp peak in $\chieff$. 
Instead, we have employed a hybrid approach that allows us to model all significant correlations in the population without increasing the dimensionality of \PixelPop, as described below.

From Fig.~\ref{fig:MI for CE pop}, we have $\rho_I(m_1,z) \approx 0.13$ and $\rho_I(q,z) \approx 0$, which means that $m_1$ and $q$ are nearly independent of $z$. On the other hand, there is a more significant correlation between $\chi_\mathrm{eff}$ and $z$, with $\rho_I(\chi_\mathrm{eff},z) \approx 0.41$.
Therefore, we consider a ``semiparametric'' approach: we use the parametric \textsc{Powerlaw} redshift evolution model from before, but attempt to capture the $\chieff$--$z$ correlation by modeling the redshift power-law index $\kappa$ with a cubic spline, $\kappa=\varphi(\chieff)$. This allows us to encode a prior assumption that the merger rate evolves as a \pr over redshift, but still with a flexible assumption about how this evolution depends on $\chi_\mathrm{eff}$.
The spline consists of six evenly spaced nodes in the range $\chieff \in [-1,1]$. The amplitude of each node is a hyperparameter of the model and is assigned a standard normal prior. With these considerations, our merger rate is modeled as
\begin{align}
\label{eq:(3+1)d-model}
\mathcal{R}(m_1,q,\chieff;z)
=
\mathcal{R}(m_1,q,\chieff) (1+z)^{\varphi(\chieff)}
\, .
\end{align}

Hereon, we refer to Eq.~\eqref{eq:(3+1)d-model} as the \Hybrid \PixelPop model.

\begin{figure*}
\centering
\includegraphics[width=0.32\textwidth]{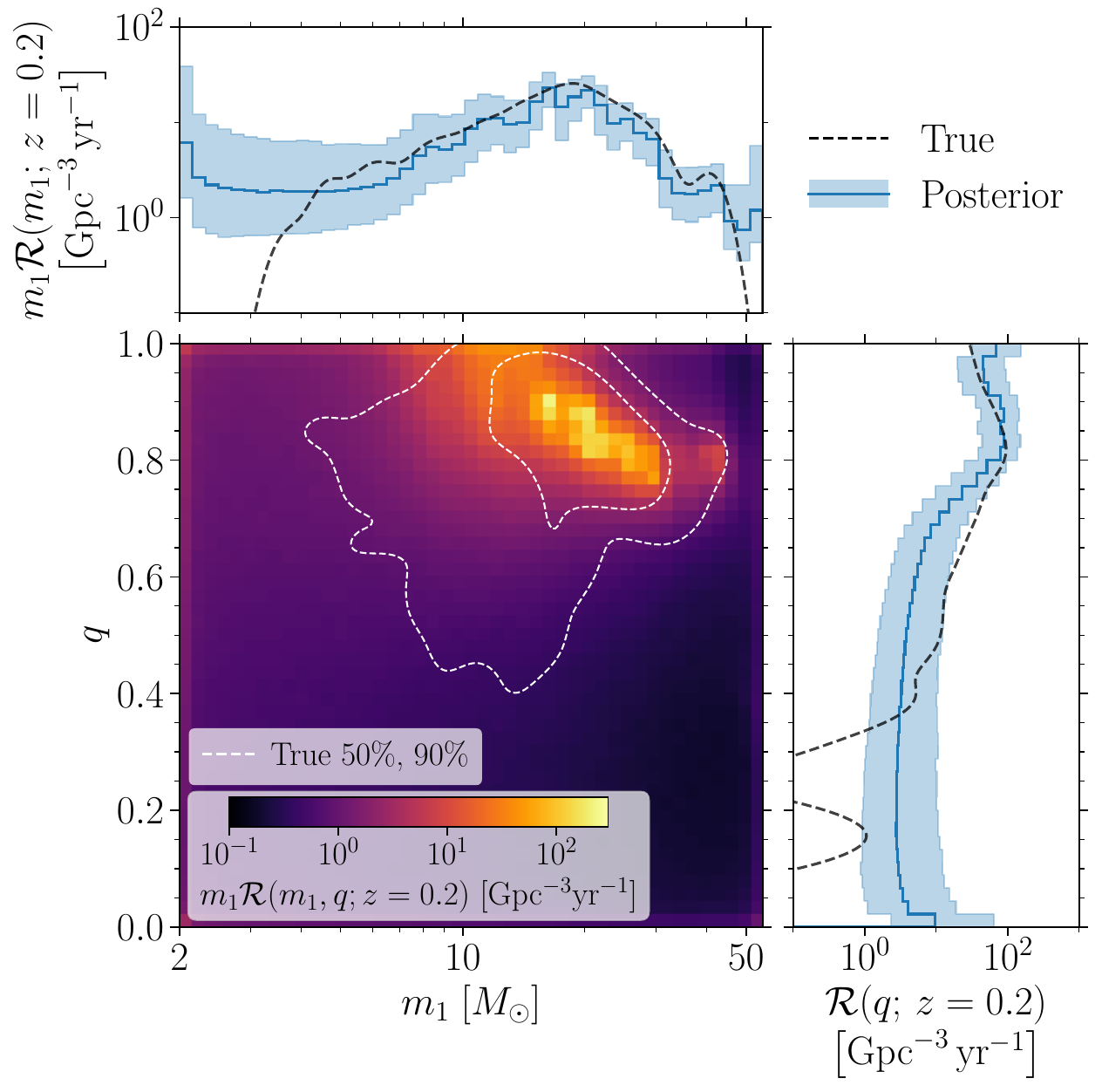}
\includegraphics[width=0.32\textwidth]{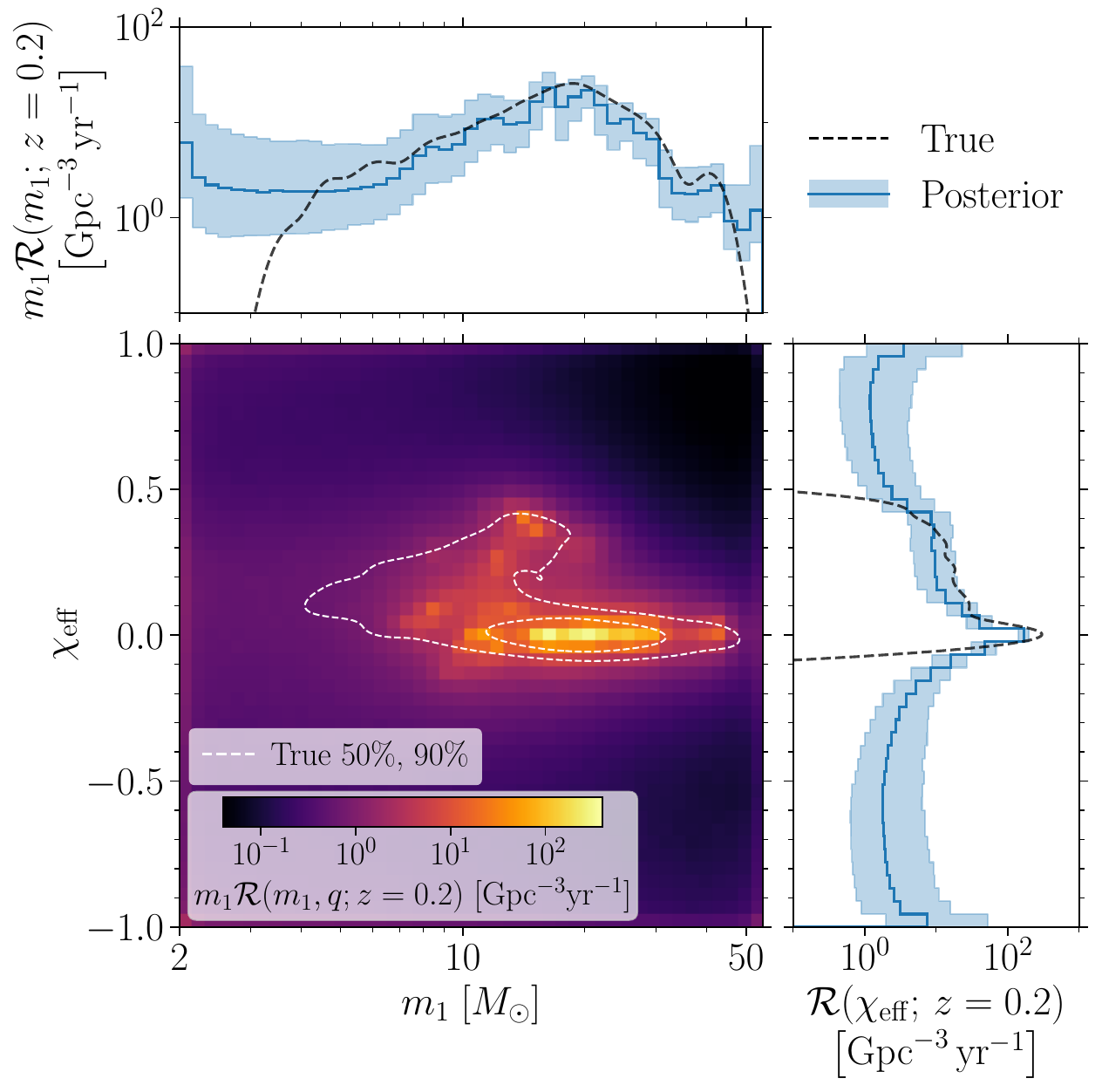}
\includegraphics[width=0.32\textwidth]{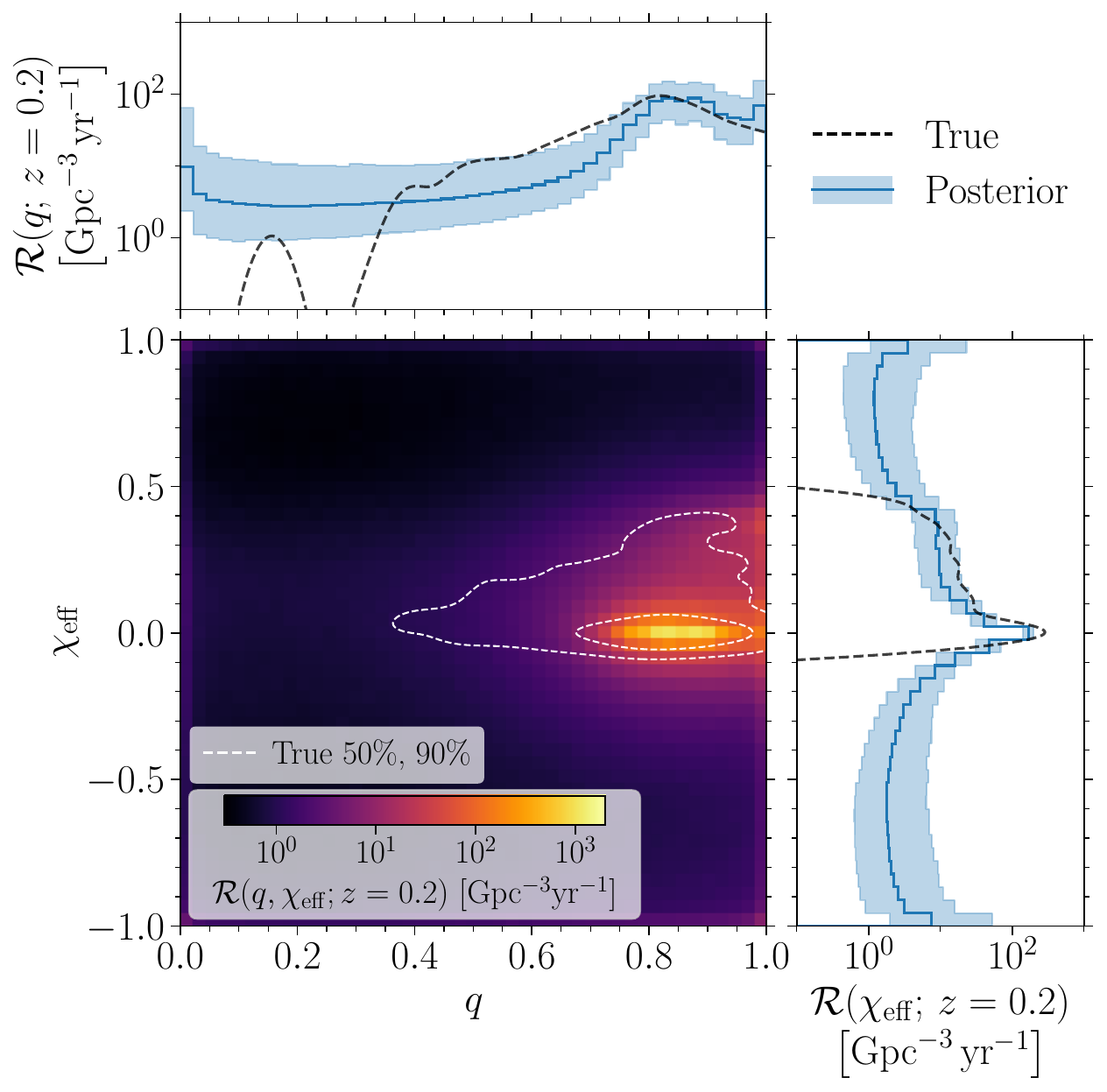}
\includegraphics[width=0.32\textwidth]{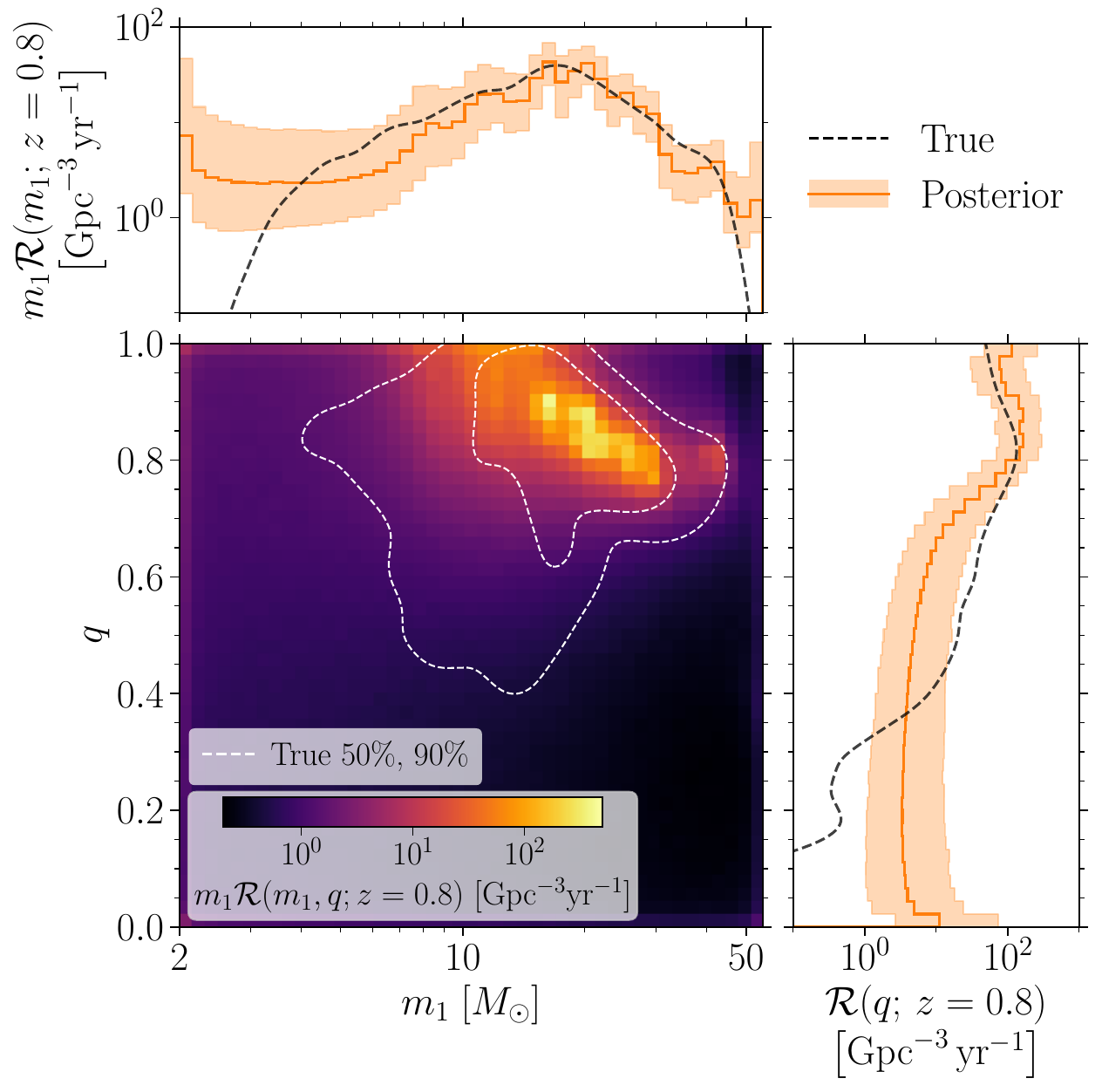}
\includegraphics[width=0.32\textwidth]{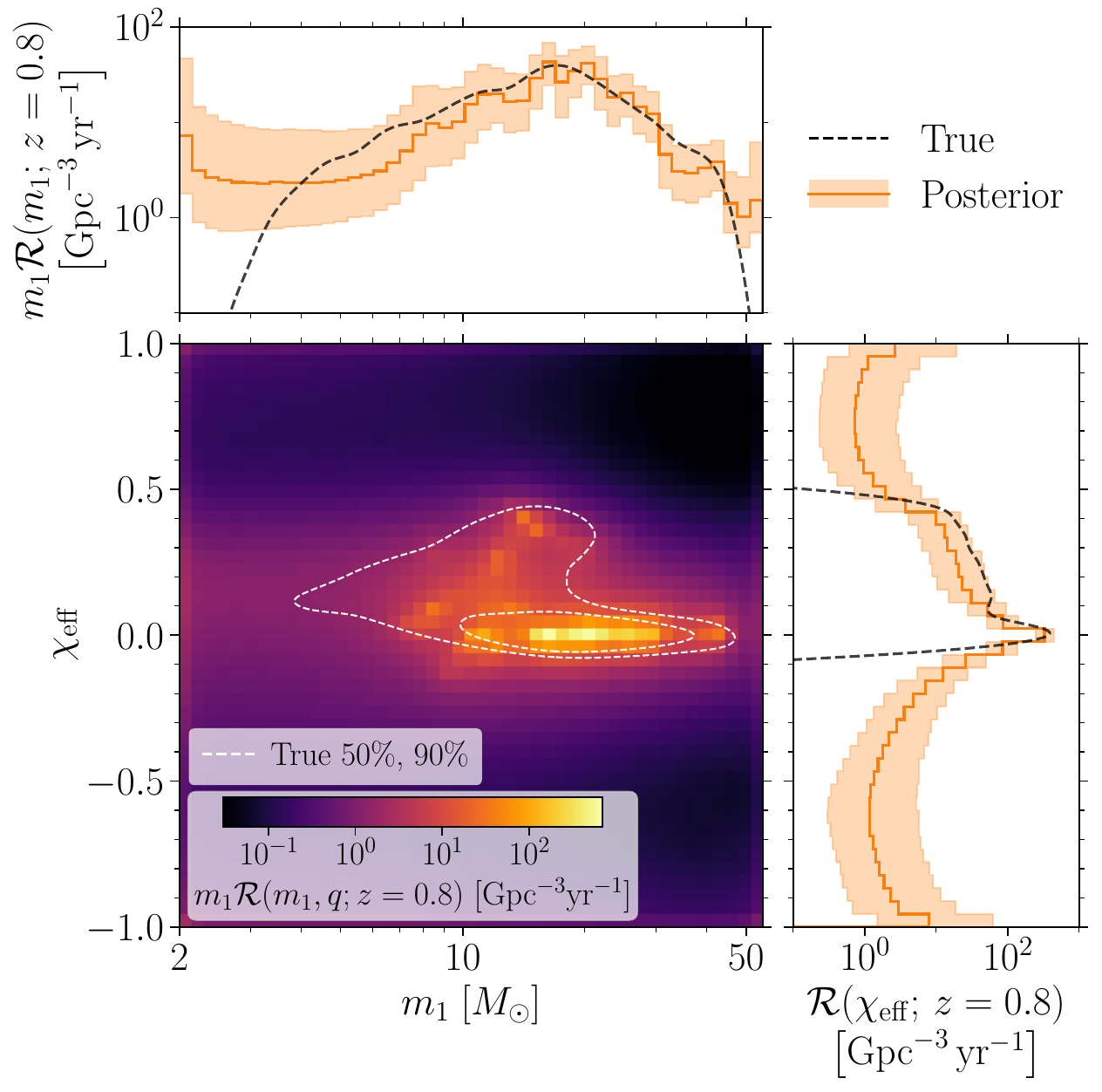}
\includegraphics[width=0.32\textwidth]{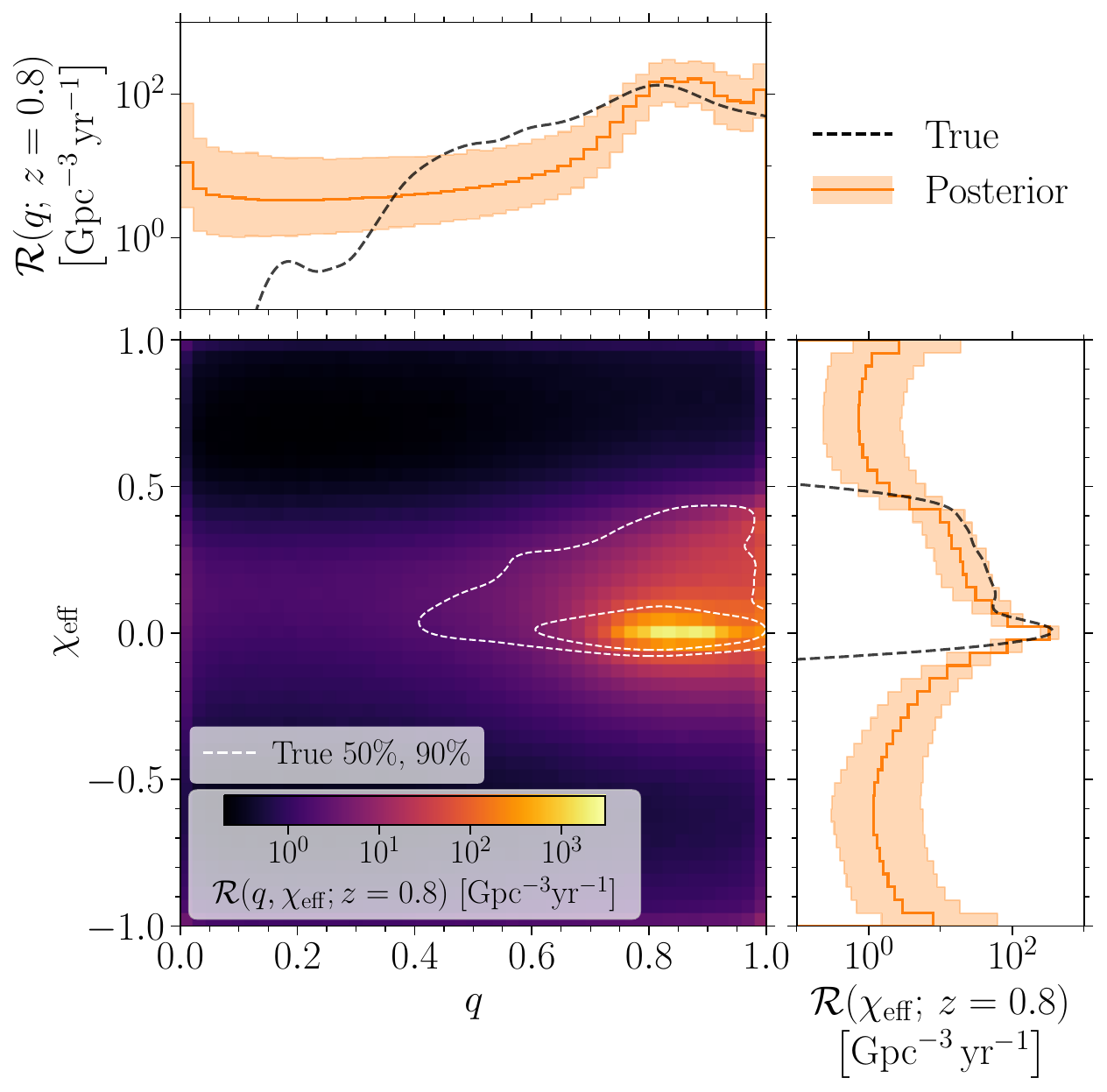}
\caption{
Similar to Fig.~\ref{fig:3d}, but for our \Hybrid model: the inferred comoving merger-rate densities $\mathcal{R}(m_1,q;z)$ (left), $\mathcal{R}(m_1,\chieff;z)$ (center), and $\mathcal{R}(q,\chieff;z)$ (right), evaluated at redshifts $z=\zone$ (top) and $z=\ztwo$ (bottom). In each figure, the central panel shows the median of the two-dimensional merger-rate posterior, along with 50\% and 90\% credible regions of our true underlying population; brighter regions indicate a higher inferred merger rate. Similarly, each upper (right-hand) panel shows the rate marginalized over the source parameter on the vertical (horizontal) axis; the dashed line represents the true population, while the solid line and shaded region represent the posterior median and central 90\% credible region.
}
\label{fig:3dcond}
\end{figure*}

Figure~\ref{fig:3dcond} shows the inferred merger rate evaluated at $z=0.2$ (top row) and $z=0.8$ (bottom row). Like before, the one- and two-dimensional marginals over primary mass and mass ratio are consistent with both the true distributions and the inferred rates using the two- and three-dimensional \PixelPop models, aside from small deviations not attributable to model systematics. At both redshift values, the true mass-ratio distribution in the range $0.5 \lesssim q \lesssim 0.7$ is again slightly above the central 90\% credible region. In fact, at $q\approx 0.65$, the truth lies at the $98\%$ level of the inferred rate at both $z=0.2$ and $z=0.8$. However, similar to three-dimensional \PixelPop, we do not find these deviations between the true and inferred rates with our catalog of point-estimate events (see discussion in App.~\ref{app:appDeltas}).

The minimum and maximum 90\% relative rate uncertainties we recover for $m_1$ and $q$ are shown in the third row of Table~\ref{tab:relrates}. For the effective spin, we get similar relative uncertainties to those with three-dimensional \PixelPop. The smallest (largest) $\relrate$ is 0.4 (1.5) at $\chieff\approx0$ ($\chieff\approx0.3$). Additionally, we still find a small deviation from the truth at the peak of the distribution: the true rate lies at the 99\% level of the inferred rate. Again, this is due to the imperfect approximation of the likelihood for narrow populations (see App.~\ref{app:appDeltas} for further discussion).

The most significant result for the \Hybrid \PixelPop model with respect to the two- and three-dimensional \PixelPop-only analyses is for the marginal effective spin distribution at $z=\ztwo$. As shown in the bottom-right panel of Fig.~\ref{fig:3d-comparison}, the inferred $\mathcal{R}(\chieff;z=\ztwo)$ is now within the 90\% credible interval for $0 \leq \chieff \leq 0.5$. We now find the true peak of the distribution at the $61\%$ level of the inferred rate (compare to $100\%$ in Sec.~\ref{sec:2d-inference} and $87\%$ in Sec.~\ref{sec:3d-inference}). Additionally, at $\chieff\approx0.2$---the point of maximum deviation between the inferred and true rates in the earlier analyses---we now find that the truth lies at the $92\%$ level of the inferred posterior distribution. This represents a significant improvement from three-dimensional \PixelPop at $z=\ztwo$, shown in the bottom-left panel of Fig.~\ref{fig:3d-comparison}, where we observed a considerable bias due to the unmodeled $\chieff$--$z$ correlation. We further analyze this correlation in App.~\ref{app:xeff-z} by evaluating $\mathcal{R}(\chieff;z)$ at different values of the effective spin. 

With the \Hybrid \PixelPop model, we recover the parameter distributions within the 90\% posterior credible intervals, except for deviations from the truth not attributable to model systematics (see App.~\ref{app:appDeltas}). These results highlight that all significant correlations (for the \ac{CE} population, all except the $m_1$--$z$ and $q$--$z$ correlations) must be modeled to recover the merger rate across the parameter space with minimal bias. \PixelPop captures these correlations with minimal assumptions, without imposing strong functional forms or specifying a priori which parameters are correlated, if any. In our analysis, we use knowledge of the true underlying population only to decide which parameters \emph{may} be correlated. When analyzing real \ac{GW} data, one does not have access to the real distribution of population parameters. We discuss the implications for analyses of real data in Sec.~\ref{sec:conclusions}.

\section{Are formation channels distinguishable?}
\label{sec:cross-entropy}

The results we presented above show that \PixelPop is a promising tool to measure the merger rate across a complicated \textit{multidimensional} parameter space. However, this unmodeled inference alone does not provide information about the separate astrophysical formation channels that may contribute to the overall merger rate. In this section, we introduce a method
to determine whether we can distinguish between different formation channels using the results from \PixelPop. 

Given a set of $M$ synthetic sources whose true parameters $\{\theta_i\}_{i=1}^M$  are drawn from an astrophysical distribution $q(\theta)$, e.g., from a population-synthesis simulation (in our case, the \ac{CE} population), the likelihood that they were instead independent draws from a different population model $p(\theta)$ is $p (\{\theta_i\}_{i=1}^M) = \prod_{i=1}^M p(\theta_i)$. The \textit{similarity} of these two populations is encoded in the shape of their distributions across the parameter space of $\theta$, not the number of simulated sources $M$. Therefore, we quantify the similarity $\mathcal{S}(p,q)$ with the average log-likelihood

\begin{align}
\label{eq: average likelihood}
\mathcal{S}(p,q) 
=
\frac{1}{M} \sum_{i=1}^M \log p (\theta_i)
\approx
\int \dd{\theta} q(\theta) \log p(\theta)
\, .
\end{align}

To assess how well a proposed astrophysical population $q$ matches an observed catalog of \ac{GW} events, we use \PixelPop to first encode our knowledge of the underlying population from which those events came without making strong assumptions.
Specifically, we use the \Hybrid \PixelPop model for the reference distribution $p(\theta) \propto \dd{N}/\dd{\theta}$, where $\theta = (m_1, q, \chieff, z)$. Following Eq.~\eqref{eq:source-frame-rate}, this is
\begin{align}
\label{eq: prob density}
p(m_1,q,\chieff,z)
\propto
\frac{1}{1+z} \dv{V_\mathrm{c}}{z} \mathcal{R}(m_1,q,\chieff;z) 
\, ,
\end{align}
with an appropriate normalization constant and where $\mathcal{R}(m_1,q,\chieff;z)$ is the model
defined in Eq.~\eqref{eq:(3+1)d-model}. From the population analysis with the \Hybrid \PixelPop model, we have many posterior draws that characterize the uncertainty in the merger rate, i.e., many different realizations of the model $p(m_1, q, \chieff, z)$. We compute Eq.~\eqref{eq: average likelihood} for each of those posterior draws, thus constructing a posterior distribution for the population similarity. For two distinct astrophysical populations, their similarities to the \Hybrid \PixelPop model are correlated because they are computed using the same posterior draws. We can then use this similarity metric to assess whether the \PixelPop inference results prefer the \ac{CE} channel---the actual population that we used to produce our \ac{GW} catalog---over other channels.

\begin{figure}
\centering
\includegraphics[width=\linewidth]{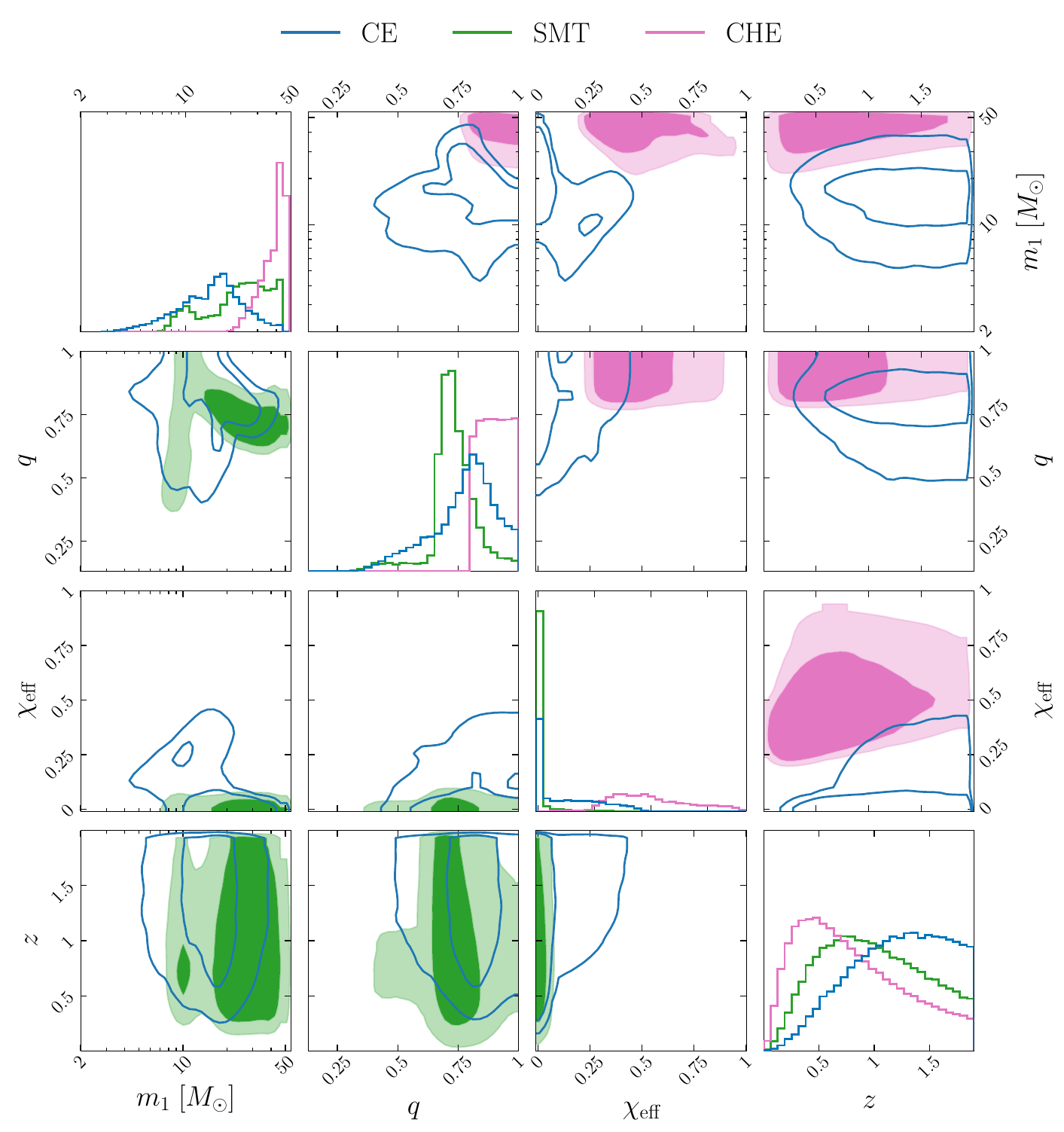}
\caption{
One- and two-dimensional marginals of the source-frame primary mass $m_1$, mass ratio $q$, effective spin $\chieff$, and redshift $z$ of the simulated \ac{CE} (blue), \ac{SMT} (green), and \ac{CHE} (pink) populations. The contours show the 50\% and 90\% credible regions. 
}
\label{fig:other-corners}
\end{figure}

In particular, we compare the \ac{CE} channel to two other isolated-binary populations simulated by \citet{Zevin:2020gbd} and introduced in Sec.~\ref{sec:introduction}: \ac{SMT} and \ac{CHE}. Similar to Fig.~\ref{fig:true_pop_corner_CE}, in Fig.~\ref{fig:other-corners} we show the populations of \ac{SMT} and \ac{CHE} sources, in comparison to the \ac{CE} population. Though these are all \ac{BBH} isolated evolution channels, the astrophysical properties of the progenitors in \ac{CHE} binaries (e.g., orbital periods $\lesssim 4$ days~\citep{Bavera:2020uch}) are significantly different from those of binaries undergoing \ac{CE} or \ac{SMT} evolution, which share several initial evolutionary stages. The key difference between these two channels is whether the mass transfer episode occurring after the first \ac{BH} is born is stable or unstable~\citep{Zevin:2020gbd, Ivanova:2012vx}. In the simulated populations we consider, binaries in the \ac{SMT} channel experience only stable mass-transfer episodes, whereas those in the \ac{CE} channel undergo a late \ac{CE} phase.

\begin{figure}
\centering
\includegraphics[width=\linewidth]{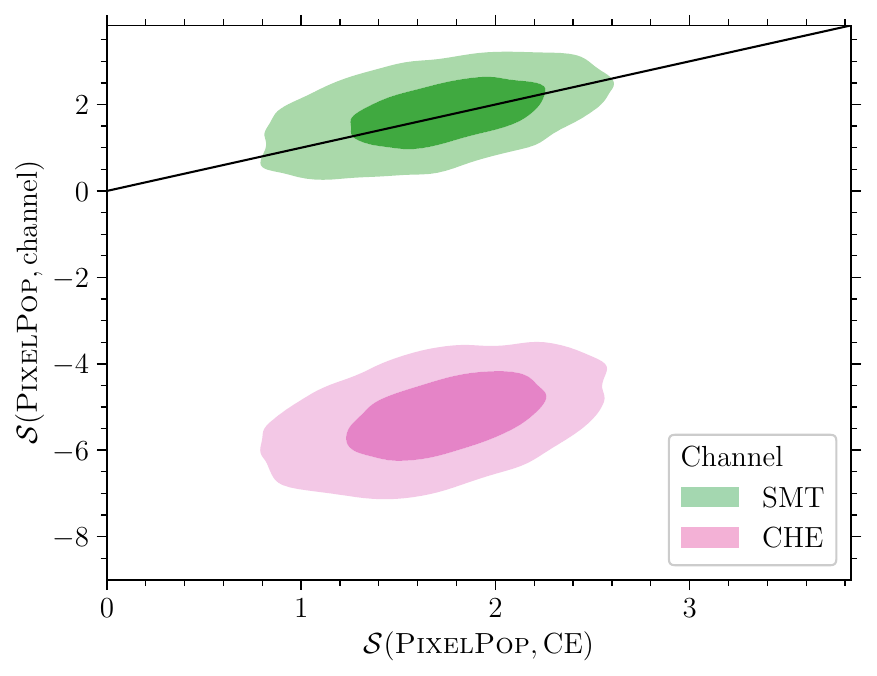}
\caption{Similarity of the simulated SMT (green) and CHE (pink) populations to the \Hybrid model (vertical axis), compared to that of the true CE population (horizontal axis). Contours show the 50\% and 90\% credible regions over the similarities computed using posterior samples for the \Hybrid population model. In regions below the diagonal line, the CE population is correctly concluded to be the true population.
}
\label{fig:cross-ent-CHE-SMT}
\end{figure}

In Fig.~\ref{fig:cross-ent-CHE-SMT}, we compare the posterior distributions of the population similarity for the \ac{CE} population with those for the \ac{SMT} and \ac{CHE} populations.
In regions below the diagonal, the CE population is more similar to the \PixelPop inference and is favored over the alternative population considered. In contrast, regions above the diagonal indicate that a \ac{CE} population is disfavored.

Compared to the CHE population, the true CE population is preferred with 100\% credibility, meaning we can tell apart a CE-only from a CHE-only population, assuming their parameter distributions are as predicted by the population synthesis simulations from~\citet{Zevin:2020gbd}. Therefore, with an O4-like catalog of 400 detections, we can use Bayesian nonparametrics to distinguish certain GW formation channels. This conclusion is not too surprising, as these two formation channels produce very different source distributions due to substantially different astrophysical processes, as seen in Fig.~\ref{fig:other-corners}. Here, we see that the population of \ac{CHE} sources peaks at $m_1 \approx 44\,\Msun$, is uniformly distributed in mass ratio for $0.8 \leq q < 1$, has positive effective spins with no sharp features, and has a steeper redshift evolution, peaking at $z\approx0.5$. In contrast, the \ac{CE} population peaks at $m_1\approx19\,M_\odot$, $q\approx0.8$, and $z>1$, and has a prominent peak at $\chieff\approx0$. Since \PixelPop analyzes joint distributions and multidimensional correlations, it can correctly identify the differences between both populations, allowing us to conclude that \ac{CE} is correctly preferred over \ac{CHE} as the formation channel. We stress that the \ac{CE} and \ac{CHE} populations are distinguishable under the assumptions of~\citet{Zevin:2020gbd}; the true channels in nature may produce different distributions.

On the other hand, when we consider the simulated \ac{SMT} population, we find that \ac{CE} is favored over \ac{SMT} at only 45\% posterior credibility. In Fig.~\ref{fig:cross-ent-CHE-SMT}, this corresponds to more than half of posterior samples lying above the equal-similarity diagonal. This means that, at O4 sensitivity with a catalog of 400 \ac{CE} sources, we cannot distinguish if the sources in the population evolved via \ac{CE} or \ac{SMT} alone, under the population synthesis simulations from~\citet{Zevin:2020gbd}. We attribute this result to \ac{CE} and \ac{SMT} being isolated binary evolution channels that share many evolutionary steps, even though they differ in some mass-transfer phases. Note from Fig.~\ref{fig:other-corners} that the mass-ratio distribution of the \ac{SMT} population peaks at $q\approx0.7$ (cf. $q\approx0.8$ in the \ac{CE} population) and it also has a sharp peak at $\chieff\approx0$. The SMT $m_1$--$q$ correlation also resembles that of the \ac{CE} population, along with the $m_1$--$\chieff$ and $q$--$\chieff$ correlations. The indistinguishability between formation channels sharing several evolutionary steps using Bayesian nonparametrics highlights the difficulty of extracting astrophysical insights from these models.

\section{Conclusions}
\label{sec:conclusions}

In this paper, we used \PixelPop~\citep{Heinzel:2024jlc, Heinzel:2024hva} to infer multidimensional correlations across the masses, spins, and redshifts of \ac{BBH} mergers from a
realistic astrophysical population of sources that evolved via \ac{CE}. We considered a catalog of $N_\mathrm{obs}=400$ sources at \ac{O4} sensitivity with a three-detector network of LIGO and Virgo. Previous \PixelPop analyses have focused on two-dimensional correlations only~\citep{Heinzel:2024hva,Heinzel:2024jlc}. However, due to the complex astrophysical processes governing \ac{CE} evolution, the simulated population we considered here exhibits nontrivial and nonlinear correlations across the four-dimensional space of primary \ac{BH} mass $m_1$, binary mass ratio $q$, effective spin $\chi_\mathrm{eff}$, and merger redshift $z$.

We employed the \acf{MI} to quantify the degree of correlation between source properties in the population, finding that the most correlated pairs in our simulated population are (in descending order): $(m_1,q)$, $(m_1,\chieff)$, $(q,\chieff)$, and $(\chieff,z)$.
Therefore, we followed a multi-step approach to infer the underlying population and its correlations, using \PixelPop to prevent imposing strong \textit{a-priori} assumptions. By progressively allowing for population-level correlations in our analyses, we illustrate the biases that can arise from using insufficient models: (1) we modeled the $m_1$--$q$ correlation with \PixelPop, but used univariate parametric models for $\chi_\mathrm{eff}$ and $z$, which led to significant biases due to the unmodeled correlations involving effective spin (Sec.~\ref{sec:2d-inference}); (2) we included the correlations between $\chieff$ and mass parameters, but still found deviations from the truth for the inferred effective spin distribution at higher redshift due to the significant---yet unmodeled---evolution of $\chieff$ with $z$ (Sec.~\ref{sec:3d-inference}); (3) finally, besides modeling $(m_1,q,\chieff)$ with \PixelPop, we used a hybrid approach to account for the correlation between $\chi_\mathrm{eff}$ and $z$ by taking advantage of the near independence of $m_1$ and $q$ with redshift (Sec.~\ref{sec:(3+1)d-inference}). 

With the \Hybrid model, we successfully recover the astrophysical merger rate in all source parameters with no bias, aside from small deviations from the 90\% credible region not attributable to model systematics (see App.~\ref{app:appDeltas} for further discussion). \textbf{These results highlight \textsc{PixelPop's} capability of inferring population-level correlations for large catalogs of realistic sources.} However, in pursuing our multi-step approach, we utilized our knowledge of the true population. With real \ac{GW} observations, we cannot \textit{a-priori} ignore any (potential) correlations in the underlying population. The posteriors for our three models all passed one-dimensional posterior predictive checks, even though we know the two- and three-dimensional \PixelPop models are not correctly inferring the underlying population. Indeed, previous work has shown that standard posterior predictive checks often fail \citep{Romero-Shaw:2022ctb, Miller:2024sui}, and we will therefore pursue more robust checks that account for correlations in future work.

Having a model that accounts for all significant correlations with no bias allowed us to test whether meaningful astrophysical insights can be extracted from the inference results. While nonparametric models generally lack interpretability, \textbf{we devised a model-independent method to compare the \textit{similarity} of proposed astrophysical populations with the nonparametric results from \PixelPop. With current sensitivities and near-future catalogs, we found that only channels with significantly different astrophysical processes can be distinguished.}

In particular, considering a \ac{BBH} population in which all binaries undergo a late \ac{CE} phase, we found that \ac{CE} is correctly favored as the dominant evolutionary mechanism over \ac{CHE} due to the distinct correlations imprinted by both formation channels. On the other hand, we found that it is not possible to distinguish whether the \acp{BBH} in the underlying population evolved via \ac{CE} or \ac{SMT} alone. This is because both channels share several evolutionary stages, which imprint similar features in the multidimensional correlations, making them difficult to tell apart with a nonparametric model. In this way, similar formation channels remain undistinguishable under nonparametric analyses at current sensitivity with near-future catalogs. Our results contrast sharply with those from astrophysically informed population models~\citep{Zevin:2020gbd, Colloms:2025hib}: while these can confidently infer the branching fractions of formation channels in \ac{BBH} populations, they also rely on strong assumptions that we cannot make in reality, which can lead to highly biased results (See e.g.,~\citet{Cheng:2023ddt}). Since we cannot distinguish between populations where all sources evolve via one particular channel, we have even less chance of probing mixtures of populations in a nonparametric way, which is why we did not consider true populations consisting of a mixture of different formation channels. In future work, we plan to extend the similarity metric to account for populations originating from multiple formation channels. Such an extension could be applied to real \ac{GW} catalogs, for which the underlying \ac{BBH} population likely contains contributions from several astrophysical pathways~\citep[e.g.,][]{Mapelli:2021taw,Zevin:2020gbd,Mandel:2018hfr,Cheng:2023ddt,Colloms:2025hib}.

With the anticipated increase in sensitivity of ground-based \ac{GW} detectors over the next few years, \ac{BBH} catalogs are expected to expand well beyond the datasets considered in this paper. By the conclusion of the fifth observing run, there may be more than 1000 \ac{BBH} observations~\citep{Kiendrebeogo:2023hzf}. Such a substantial dataset may allow us to resolve population-level differences that are currently elusive, as we showed in this paper. For example, the \ac{CE} and \ac{SMT} formation channels---which we found to be indistinguishable with a sample of 400 \acp{BBH}---may become distinguishable as subtle statistical features emerge with a larger number of sources. To maximize our ability to uncover these features, the development of algorithms capable of inferring the high-dimensional structure of the parameter space is crucial. \PixelPop represents an important step in that direction.

\section*{acknowledgments}
We thank Cailin Plunkett, Thomas Dent, Yin-Jie Li, and Noah Wolfe for useful discussions. We also thank Storm Colloms as our internal LIGO reviewer.
S.A.-L. is supported by the Mario Santo Domingo and Thomas A. Frank fellowship funds at MIT.
J.H. is supported by the NSF Graduate Research Fellowship under Grant No. DGE1122374.
M.M. is supported by the LIGO Laboratory through the National Science Foundation awards PHY-1764464 and PHY-2309200 and a Research Fellowship from the Royal Commission for the Exhibition of 1851.
J.H. and S.V. are partially supported by the NSF grant PHY-2045740.
The authors are grateful for computational resources provided by the LIGO Laboratory and supported by National Science Foundation Grants PHY-0757058 and PHY-0823459.

\clearpage

\appendix
\section{Sensitivity injections}\label{app:vtInjs}

To account for selection effects when recovering the \ac{CE} population, we perform a simulation campaign to characterize the sensitivity of the O4 LIGO--Virgo network to \acp{BBH} spanning a range of parameters. We draw the \ac{BBH} injections from the following distributions: a \textsc{Power Law} redshift model~\citep{Fishbach:2018edt}, a mixture of two Gaussian distributions for the effective spin: one narrow for the peak at $\chieff\approx0$ and one wide for the bulk of the distribution. We use a normalizing flow~\citep{Papamakarios:2019fms, Kobyzev:2019ydm} implemented with \textsc{FlowJAX}~\citep{ward2023flowjax} to approximate the two-dimensional $m_1$--$q$ distribution and draw 80\% of the injections from it. Our normalizing flow consists of only three coupling layers with affine transformations~\citep{dinh2016density}, each using a conditioner network with two hidden layers of 10 ReLU-activated neurons. We sample the remaining 20\% of the sources from uniform joint distributions in $\log{m_1}$ and $q$. For all the other parameters, we sample from the same priors as used for \ac{PE}. We draw a total of $8.5 \times 10^8$ sources, $2 \times 10^6$ of which are above our SNR threshold of $\rho_\mathrm{opt}=11$, as in Sec.~\ref{sec:detectable catalog}.

\section{The intrinsic conditional autoregressive (ICAR) model}
\label{sec:appICAR}

In our original implementation introduced in~\citet{Heinzel:2024jlc}, we used a \ac{CAR} prior for the binned multi-dimensional rate density. The \ac{CAR} prior is given as \citep{Besag:1974abc, Besag:1995car}:
\begin{align}
& p ( \ln\boldsymbol{\mathcal{R}} | \kappa , \sigma , \mu )
=
\sqrt{
\frac
{ \det ( \mathbf{D} - \kappa \mathbf{A} ) }
{ (2\pi \sigma^2 )^B }
}
\nonumber \\
& \times
\mathrm{exp}
\bigg[
- ( \ln\boldsymbol{\mathcal{R}} - \mu \mathbf{I} )^\mathrm{T}
\frac { \mathbf{D} - \kappa \mathbf{A} } { 2 \sigma^2 }
( \ln\boldsymbol{\mathcal{R}} - \mu \mathbf{I} )
\bigg]
\, ,
\label{eq: car prior}
\end{align}
where: $\boldsymbol{\mathcal{R}}$ is a vector representation of all the multidimensional binned rates $\{\mathcal{R}_b\}_{b=1}^B$, $\mathbf{I}$ is a vector of ones; $\mathbf{A}$ is the adjacency matrix, where $A_{ij} = 1$ if the $i^{\rm th}$ and $j^{\rm th}$ bins are adjacent and 0 otherwise; and $\mathbf{D}$ is a diagonal matrix in which the diagonal entries are the number of adjacent sites to each bin.
The \ac{CAR} model is conditional on three parameters: $\kappa$, $\sigma$, and $\mu$. The mean $\mu$ and the coupling parameter $\sigma$ set the global scale and variation of $\boldsymbol{\mathcal{R}}$, respectively. The correlation parameter $\kappa \in [0,1]$ determines the coupling between neighboring bins. In particular, when $\kappa=1$, adjacent bins are conditionally coupled, while at $\kappa=0$ neighboring bins are independent of each other (See~\citet{Heinzel:2024jlc} for more details). The CAR model is much more efficient to evaluate than, e.g., a Gaussian process prior, because the covariance matrix determinant is trivial after evaluating the eigenspectrum of $\mathbf{D}^{-1}\mathbf{A}$, whereas the Gaussian process requires a matrix inversion and determinant at every evaluation. However, the large number of bins requires computing this eigenspectrum for a very large matrix: for models with $\mathcal{O}(10^5)$ bins total considered here, this can take days.

In this work, we instead use a \ac{CAR} model in the limit of $\kappa \to 1$, also known as an \ac{ICAR} model. Indeed, we noticed in Refs.~\citep{Heinzel:2024hva,Heinzel:2024jlc} that the posterior on $\kappa$ for the \ac{GW} population inference problem peaked strongly at 1, i.e., the \ac{GW} population is more easily explained by coupling the neighboring bins. In the \ac{ICAR} limit, the probability density becomes independent of $\mu$ and may be written
\begin{align}
& p ( \{ \ln\mathcal{R}_b \} | \sigma )
\propto
(2\pi \sigma^2 )^{-B/2}
\nonumber \\
& \times
\mathrm{exp}
\bigg[
- \sum_{i<j \; \textrm{adjacent}}
\frac{(\ln\mathcal{R}_i -\ln\mathcal{R}_j)^2}{\sigma^2}
\bigg]
\, .
\label{eq: icar prior}    
\end{align}
where the sum inside the exponential is performed over unique pairs of adjacent bins. The probability density in Eq.~\eqref{eq: icar prior} is invariant to a constant shift to the $\ln\boldsymbol{\mathcal{R}}$ vector, which implies the prior does not have finite normalization (i.e., it is an improper prior). In fact, we drop the determinant term (which normalizes the distribution), since the matrix $\mathbf{D}-\mathbf{A}$ is singular and therefore has determinant zero. The determinant $\det(\mathbf{D} - \kappa\mathbf{A})$ is only a constant in the prior, and hence we can drop this term in the prior without affecting the \PixelPop posterior for inference with fixed $\kappa$. This justifies the removal of the normalization term $\det(\mathbf{D} - \mathbf{A}) = 0$. The scale for $\ln\boldsymbol{\mathcal{R}}$ is set by the likelihood function during inference, which makes the Bayesian posterior a well-defined probability distribution with a finite normalization.

\section{Removing correlations as a sanity check}
\label{sec:appUncorrPop}

In Section~\ref{sec:2d-inference}, we modeled the most correlated parameter pair ($m_1,q$) with two-dimensional \PixelPop and we observed significant biases in the effective spin distribution recovered with the \PeakTukey parametric model. Here, we find that the model misspecifications discussed in Section~\ref{sec:2d-inference} arise from neglected correlations in the population---especially those involving $\chieff$---rather than from limitations in the assumed parametric model for the effective spin. To confirm the source of the bias, we artificially remove all correlations in the population except that between $m_1$ and $q$. We rerun full \ac{PE} and repeat the procedure of Sec.~\ref{sec:detectable catalog} to get a catalog of $N_\mathrm{obs} = 400$ detections. For inference, we again model the merger rate $\mathcal{R}(m_1,q,\chieff;z)$ as defined in Eq.~\eqref{eq: 2d rate}. 

\begin{figure}
\centering
\includegraphics[width=\columnwidth]{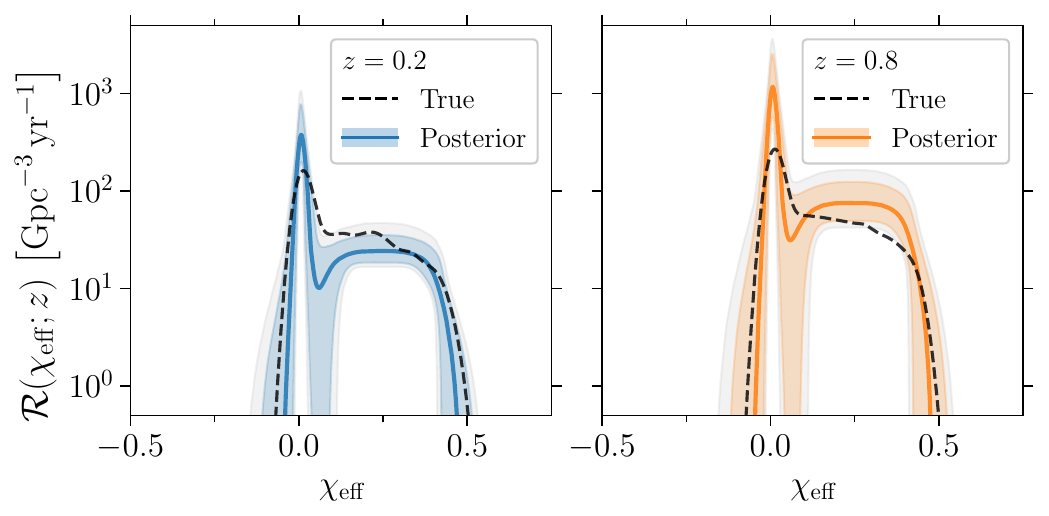}
\caption{Inferred merger-rate density $\mathcal{R}(\chieff; z)$ at redshifts $z=0.2$ (left, in blue) and $z=0.8$ (right, in orange) with the \PeakTukey model on the population with artificially removed correlations. Compared to Fig.~\ref{fig:2d-xeff-comparison}, the true distribution (dashed lines) is now within the 99\% credible interval (shaded gray region) of the inferred rate.}
\label{fig:2d-xeff-comparison-uncorr}
\end{figure}

Similar to Fig.~\ref{fig:2d-z0p2p8}, we recover the one- and two-dimensional comoving merger-rate densities for the parameters inferred with \PixelPop at both $z=\zone$ and $z=\ztwo$ within the 90\% credible regions. Figure~\ref{fig:2d-xeff-comparison-uncorr} shows the one-dimensional effective-spin distribution inferred with the \PeakTukey model at $z=0.2$ and $z=0.8$, along with 90\% and 99\% credible intervals. Compared to Fig.~\ref{fig:2d-xeff-comparison}, we now recover the true one-dimensional effective-spin distribution.

When studying the fully correlated scenario with two-dimensional \PixelPop in Sec.~\ref{sec:2d-inference}, we found that at $\chieff\approx0.2$---the effective spin value in the broad component of the distribution with maximum deviation between the inferred and true rate---the posterior was inconsistent with the truth at both $z=0.2$ and $z=0.8$. Here, we find that the truth lies at the 97\% ($z=0.2$) and 6\% ($z=0.8$) levels of the inferred rate. Therefore, ignoring the correlations between effective spin and the other parameters is the primary driver of the biased inference we saw in Section~\ref{sec:2d-inference}. 

\section{Results in the absence of single-event uncertainty}\label{app:appDeltas}

Besides performing the inference on the catalog of detectable sources described in Sec.~\ref{sec:detectable catalog}, we also consider a set of $N=400$ events with point-estimates of the true values of the source properties (i.e., in the limit of no \ac{PE} uncertainty). Our purpose with these analyses is three-fold: (a) to confirm that, in the absence of \ac{PE} uncertainty, we get tighter or comparable constraints for the inferred merger rate; (b) to highlight that the bias at $\chieff\approx0$ found in Fig.~\ref{fig:3d-comparison} is a result of the imperfect Monte Carlo approximation of the likelihood for populations with very narrow features; and (c) to show that small deviations from the 90\% credible interval in the mass-ratio distribution (see Fig.~\ref{fig:3d} and Fig.~\ref{fig:3dcond}) are not attributable to model systematics.

\begin{figure}
\centering
\includegraphics[width=\columnwidth]{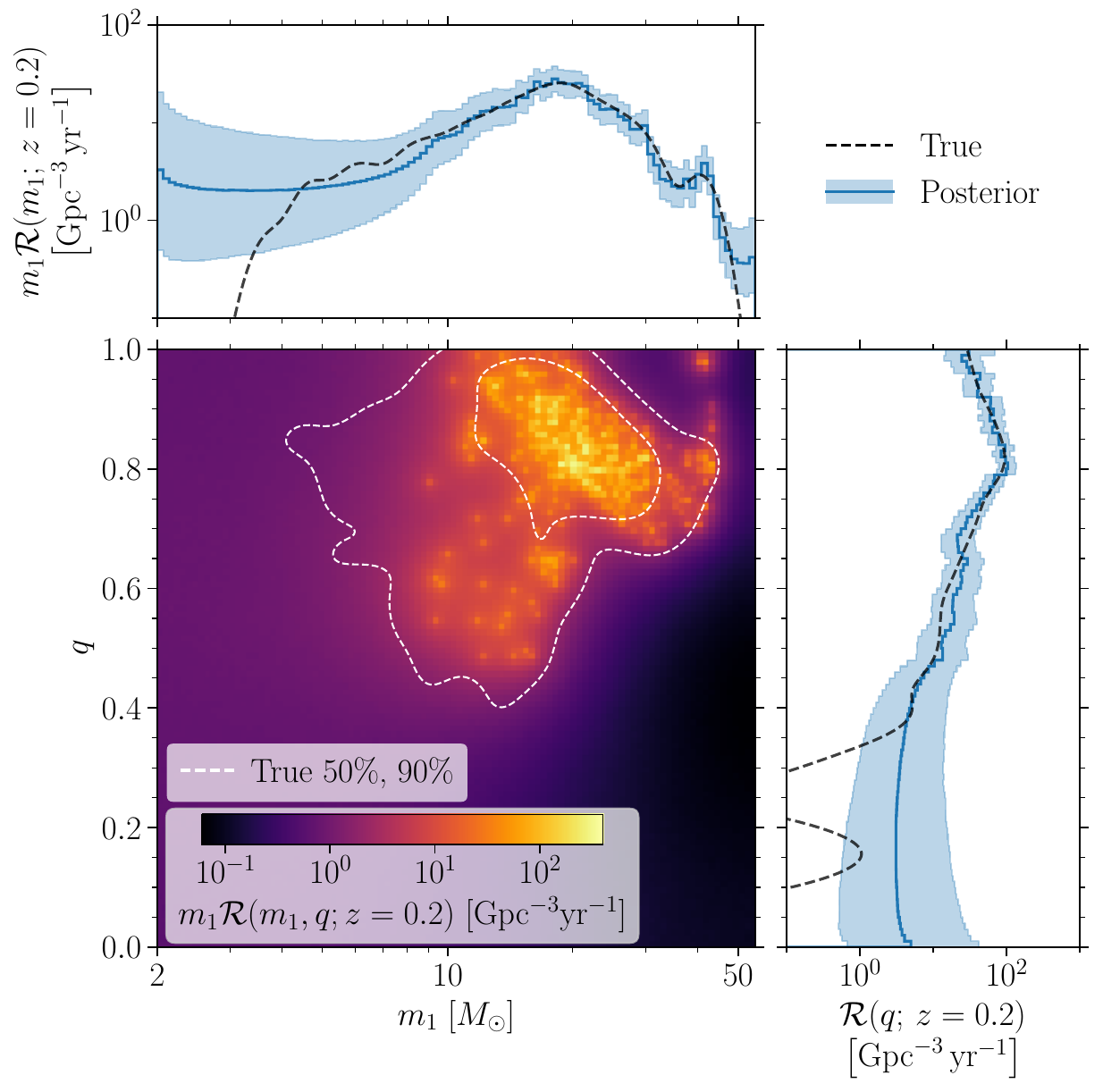}
\caption{
Inferred comoving merger-rate density $\mathcal{R}(m_1,q;z=0.2)$ for the catalog of events with point estimates for their source properties; otherwise, the same as the left panel in Fig.~\ref{fig:2d-z0p2p8}.
}
\label{fig:2d-z0p2deltas}
\end{figure}

Regarding (a), in the absence of \ac{PE} uncertainty, we recover the true distributions of all source parameters with narrower or similar credibility regions. To illustrate this, we consider the case in which we only model the $(m_1,q)$ correlation and, therefore, can achieve higher resolution in the parameter space (100 bins per dimension). Just like the left plot of Fig.~\ref{fig:2d-z0p2p8}, Fig.~\ref{fig:2d-z0p2deltas} shows $\mathcal{R}(m_1,q;z=\zone)$ (central panel), along with $\mathcal{R}(m_1;z=\zone)$ (top), and $\mathcal{R}(q;z=\zone)$ (right). The joint distribution shows finer features of the complex $m_1$--$q$ correlation. In general, we find better (comparable) constraints for the smallest (largest) relative rate uncertainty. For $m_1$, we find $\relrate=0.6$ near $m_1\approx22\,\Msun$. The largest is $3.2$ at $m_1\approx4\,\Msun$. For mass ratio, we find that the largest relative rate is $\relrate=2.8$ (near $q=0.4$) whereas the smallest is $0.6$ (at $q\approx0.8$). This is approximately half the smallest relative rate calculated in Sec.~\ref{sec:2d-inference}, as expected in the limit of no \ac{PE} uncertainty.

\begin{figure}
\centering
\includegraphics[width=\linewidth]{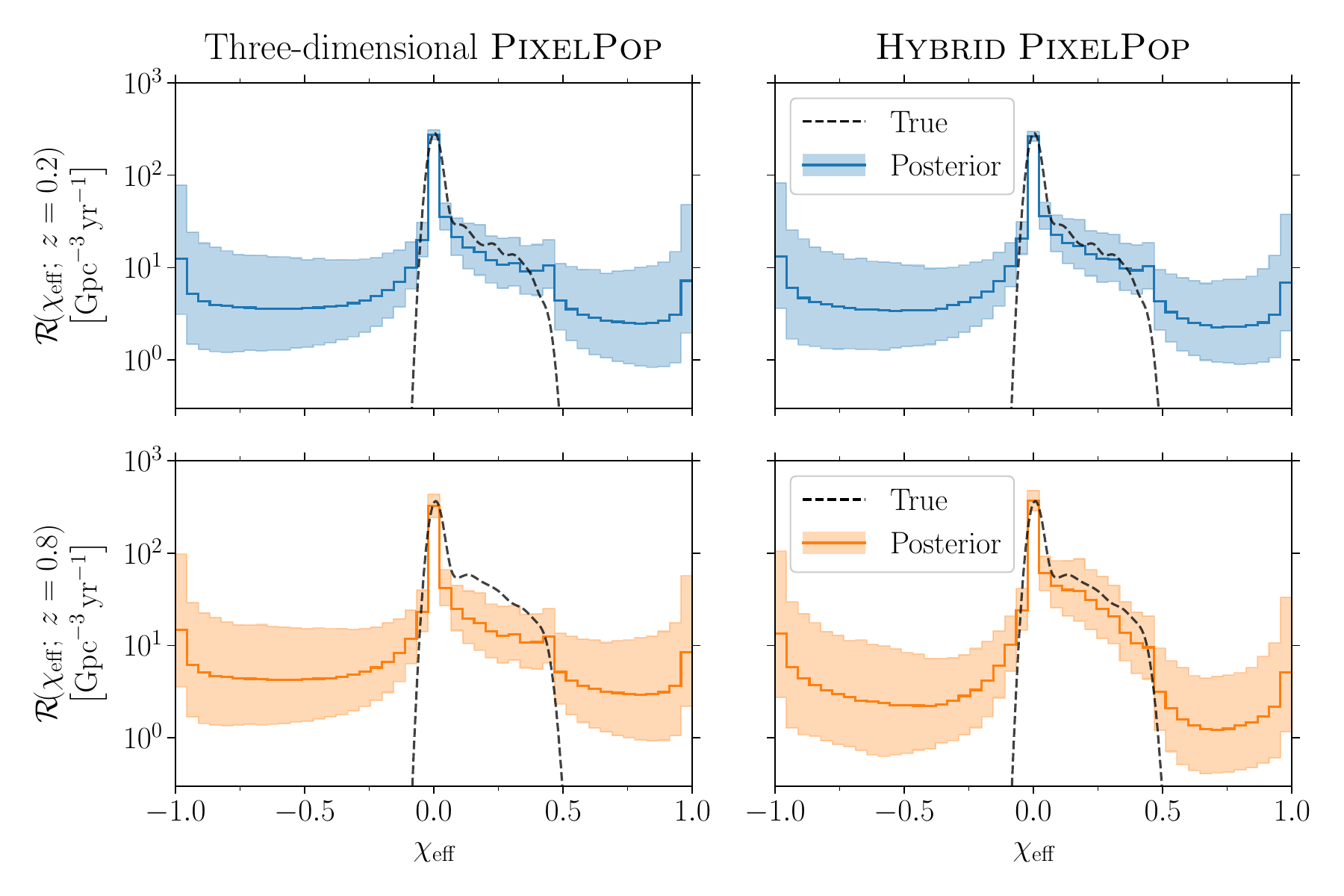}
\caption{Inferred marginalized merger-rate density $\mathcal{R}(\chieff ; z)$ at redshifts $z=0.2$ (blue) and $z=0.8$ (orange) for the three-dimensional \PixelPop (left) and \Hybrid (right) models, using the catalog of point-estimate events. Otherwise, similar to Fig.~\ref{fig:3d-comparison}. In this case, any deviation from the truth in the peak of the effective-spin distribution at $\chieff\approx\xeffone$ is minimal. However, there is significant bias at $z=0.8$ in the $0 < \chieff \leq 0.5$ region. 
}
\label{fig:3d-comparison-deltas}
\end{figure}

For (b), we confirm that the bias in the peak of the effective spin distribution seen in Fig.~\ref{fig:3d-comparison} is due to the imperfect approximation of the likelihood to narrow populations. We run three-dimensional \PixelPop and \Hybrid models using point estimates for the properties of each event in our simulated catalog. Similar to Fig.~\ref{fig:3d-comparison}, Fig.~\ref{fig:3d-comparison-deltas} shows the inferred marginalized rate $\mathcal{R}(\chieff;z)$ at $z=0.2$ and $z=0.8$, with the three-dimensional \PixelPop analysis on the left and the \Hybrid model on the right. Just like in the analysis with full \ac{PE}, at $z=0.8$ we recover the $\chieff$ peak at the 65\% and 32\% levels for three-dimensional \PixelPop and for \Hybrid \PixelPop, respectively. At $z=\zone$, the true value of the peak is now within the 90\% credible interval for both the three-dimensional \PixelPop and the \Hybrid \PixelPop model. We find that the truth lies at the 63\% and 80\% level of the inferred posterior distribution, respectively. Therefore, we see that the bias in Fig.~\ref{fig:3d-comparison} is due to the imperfect Monte Carlo approximation of the likelihood for narrow features in the population~\citep{Tiwari:2017ndi,Farr_2019}.

At $z=\ztwo$, with three-dimensional \PixelPop, we still find the bias in effective spin described in Sec.~\ref{sec:(3+1)d-inference} for the bulk of the distribution ($0 < \chieff < 0.5 $). In fact, at $\chieff\approx0.2$, where the inferred rate deviates most from the truth, the true merger rate lies at the 99\% level of the inferred merger rate posterior distribution. This confirms that we do not recover the entirety of the effective spin distribution with three-dimensional \PixelPop because we inherently neglect the $\chieff$--$z$ correlation in this analysis. 

\begin{figure}
\centering
\includegraphics[width=\linewidth]{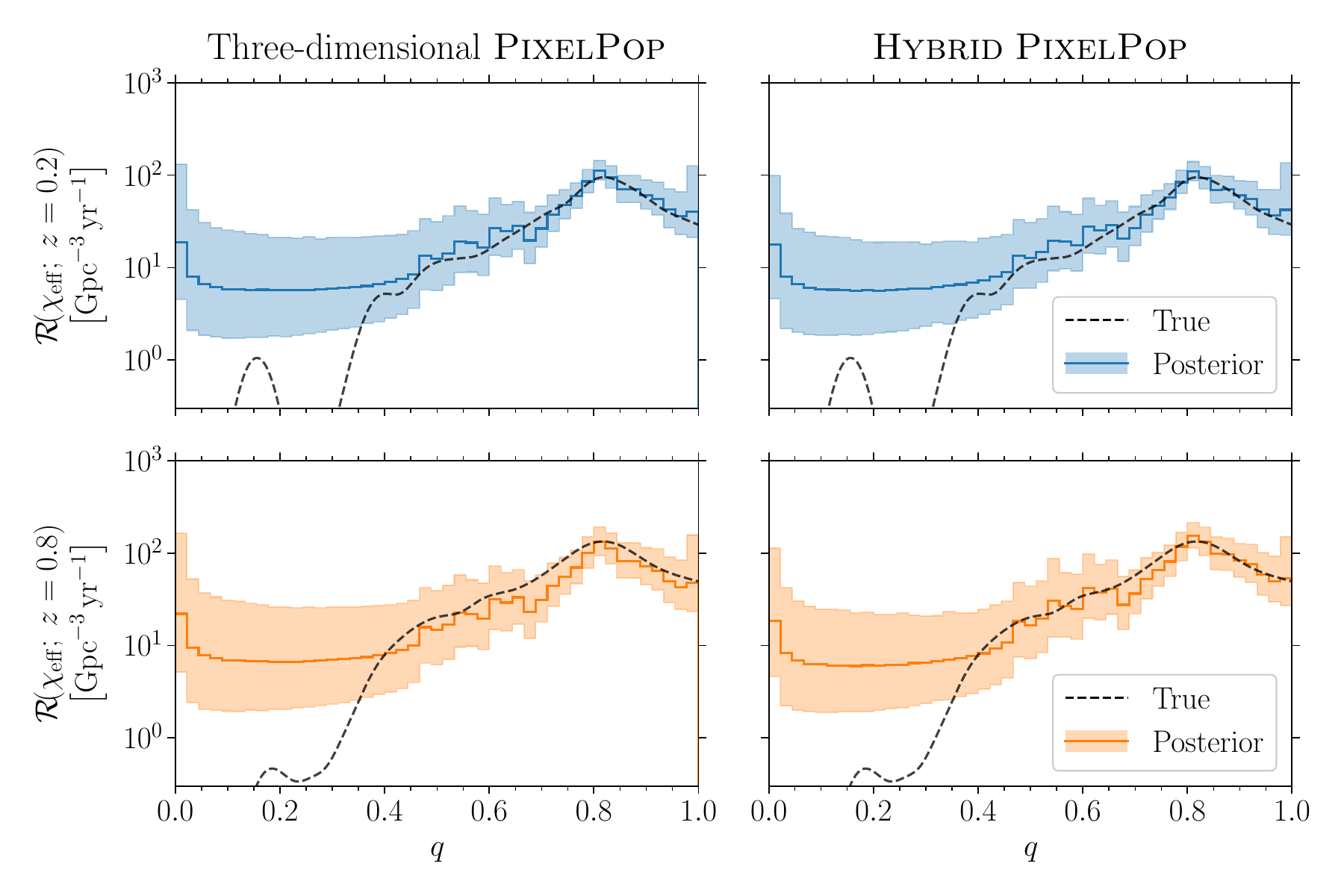}
\caption{Inferred marginalized merger-rate density $\mathcal{R}(q; z)$ at redshifts $z=0.2$ (blue) and $z=0.8$ (orange) for the three-dimensional \PixelPop (left) and \Hybrid (right) models, using the catalog of point-estimate events. The dashed line represents the true population, while the solid line and shaded region represent the posterior median and central 90\% credible region. There are no deviations between the inferred and true rate in the $0.5\lesssim q\lesssim 0.7$ region.
}
\label{fig:q-comp}
\end{figure}

Regarding (c), in Figs.~\ref{fig:3d} and \ref{fig:3dcond} we found small deviations of our inferred mass-ratio distribution from the truth at $\approx90\%$ credibility. Here, we run our three-dimensional \PixelPop and \Hybrid models on point estimates for the properties of each event in our simulated catalog. Figure~\ref{fig:q-comp} shows the inferred marginalized rate $\mathcal{R}(q;z)$ at $z=0.2$ (top row) and $z=0.8$ (bottom row) for the three-dimensional \PixelPop model (left column) and the \Hybrid model (right column) in the absence of single-event \ac{PE} uncertainty. As expected, we still find deviations from the truth at $q\lesssim0.4$ due to the absence of informative data in this region. However, in contrast with the analyses on full \ac{PE}, we now find that for both models at any redshift the true distribution in the region $q \gtrsim 0.4$ is now within the 90\% credible interval. Therefore, we confirm that the deviations in Figs.~\ref{fig:3d} and~\ref{fig:3dcond} are not attributable to model systematics.

\section{Studying the \texorpdfstring{$\chieff$}{effective spin}--\texorpdfstring{$z$}{redshift} correlation with the \Hybrid model}\label{app:xeff-z}

\begin{figure}
\centering
\includegraphics[width=\linewidth]{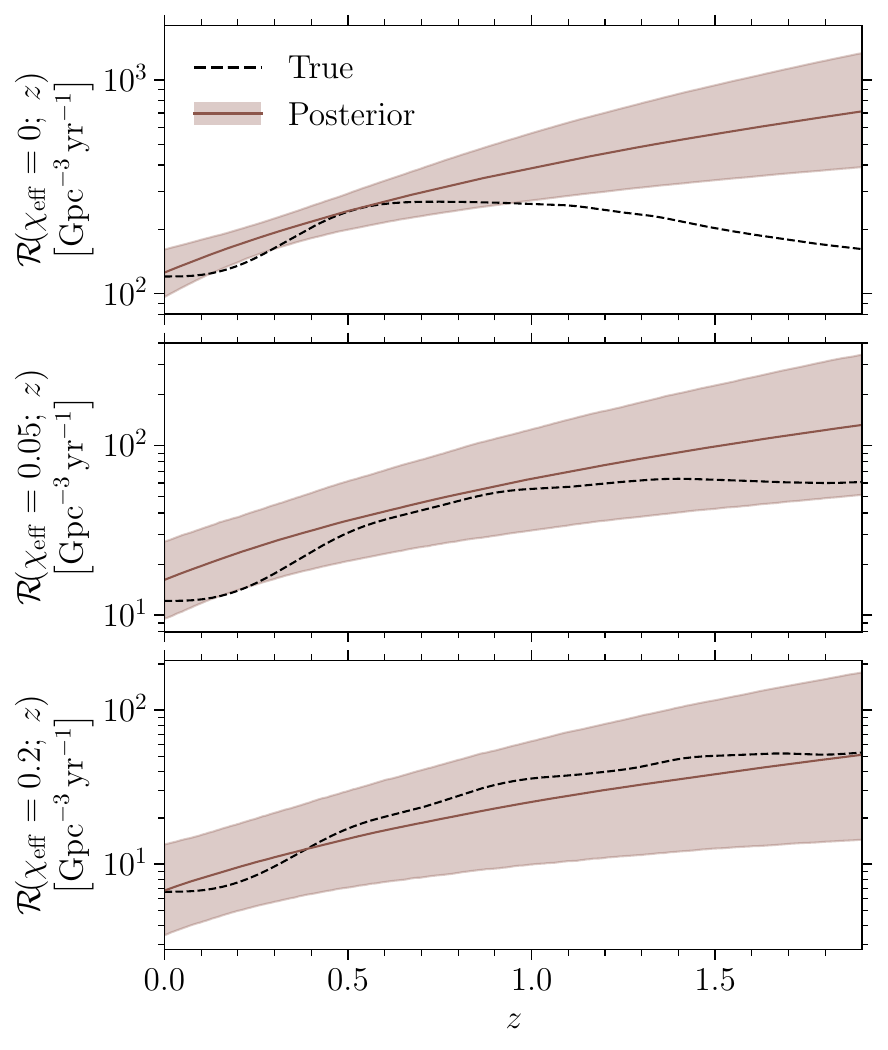}
\caption{Inferred evolution of the merger-rate density $\mathcal{R}(\chieff;z)$ over redshift and as a function of the effective spin $\chieff\approx0$ (top), 0.05 (middle), and 0.2 (bottom). The inferred median and central 90\% credible region are given by the solid line and shaded areas, respectively, while the underlying evolution of the true population is given by the dashed black line.}
\label{fig:z-at-xeff}
\end{figure}

Using the \Hybrid model defined in Eq.~\eqref{eq:(3+1)d-model}, we can study the $\chieff$--$z$ correlation in the simulated \ac{CE} population. This is the only model we can use to study such correlation because the two- and three-dimensional \PixelPop models assume independence between redshift and effective spin. Figure~\ref{fig:z-at-xeff} shows the merger-rate density $\mathcal{R}(\chieff;z)$ at $\chieff\approx0$ (the peak), $\chieff\approx0.05$ (just to the right of the peak), and $\chieff\approx0.2$ (in the broad component of the effective spin distribution). 

Note that the inferred merger-rate density $\mathcal{R}(\chieff; z)$ is consistent with the truth within the central 90\% posterior uncertainty at both $\chieff\approx0.05$ and $\chieff\approx0.2$. At these values, the flexibility of the spline that models the power-law index of the redshift evolution as a function of $\chieff$ is sufficient to capture the behavior of the true distribution. However, $\mathcal{R}(\chieff\approx0; z)$ (top panel) is only consistent within the 90\% credible interval for $z \lesssim 1$. For $z \gtrsim 1$, we see a significant deviation from the truth in the inferred merger rate evolution, which we also find when considering point estimates for the source properties of events in our catalog. The observed bias results from the redshift dependence on the effective spin evolving more steeply around the $\chieff\approx0$ peak than allowed for by our $\chieff$-dependent spline model for the redshift power-law index. There is a visible difference in the shape of $\mathcal{R}(\chieff;z)$ at $\chieff\approx0$ and $\chieff\approx0.05$. In the first case, the true merger rate decreases with redshift for $z\gtrsim 0.5$ while in the second case, the true merger rate monotonically increases with redshift. Additionally, due to the binning we use in the \Hybrid model (i.e., 45 bins for the effective spin), $\chieff\approx0$ and $\chieff\approx0.05$ lie in different bins. To better resolve the fast evolution of the redshift distribution near the effective spin peak, we aim to explore in the future the use of adaptive binning for \PixelPop. This technique would allow us to resolve with finer bins the area around the narrow $\chieff$ peak and possibly recover $\mathcal{R}(\chieff\approx0; z)$ closer to the true merger rate.

\bibliography{main}
\bibliographystyle{aasjournal}
\end{document}